\documentclass[fleqn,usenatbib]{mnras}
 
\usepackage{graphicx} 
\usepackage{txfonts} 
\usepackage{natbib}               
\title[]{Multiple populations in globular clusters and their parent galaxies\thanks{
    Based on observations with the NASA/ESA Hubble Space Telescope,
    obtained at the Space Telescope Science Institute, which is
    operated by AURA, Inc., under NASA contract NAS 5-26555.
}.}  
\author[A.\,P.\, Milone et al.] 
       {A.\,P.\,Milone$^{1}$,
       A.\,F.\,Marino$^{2}$, G.\,S.\,Da Costa$^{3}$, E.\,P.\,Lagioia$^{1}$, F.\,D'Antona$^{4}$, P.\,Goudfrooij$^{5}$, \newauthor H.\,Jerjen$^{3}$,    D.\,Massari$^{6,7}$, A.\,Renzini$^{8}$, D.\,Yong$^{3}$,
        H.\,Baumgardt$^{9}$,
        G.\,Cordoni$^{1}$,
        E.\,Dondoglio$^{1}$, \newauthor
        C.\,Li$^{10}$,  
        M.\,Tailo$^{1}$, R.\,Asa'd$^{11}$,
        E.\,M.\,Ventura$^{1}$     
\\ 
$^{1}$Dipartimento di Fisica e Astronomia ``Galileo Galilei'', Univ. di Padova, Vicolo dell'Osservatorio 3, Padova, IT-35122\\
$^{2}$ Centro di Ateneo di Studi e Attivita' Spaziali ``Giuseppe Colombo'' - CISAS, Via Venezia 15, Padova, IT-35131 \\
$^{3}$Research School of Astronomy \& Astrophysics, Australian National University, Mt Stromlo Observatory, via Cotter Rd, Weston, ACT 2611, Australia \\
$^{4}$Istituto Nazionale di Astrofisica - Osservatorio Astronomico di Roma, Via Frascati 33, I-00040 Monteporzio Catone, Roma, Italy\\
$^{5}$Space Telescope Science Institute, 3800 San Martin Drive, Baltimore,  MD 21218, USA\\
$^{6}$Kapteyn Astronomical Institute, University of Groningen, NL-9747 AD Groningen, Netherlands \\
$^{7}$Istituto Nazionale di Astrofisica - Osservatorio Astronomico di Padova, Vicolo dell'Osservatorio 5, Padova, IT-35122\\
$^{8}$ Dipartimento di Fisica e Astronomia, Universita' degli Studi di Bologna, Via Gobetti 93/2, I-40129 Bologna, Italy\\
$^{9}$ School of Mathematics and Physics, The University of Queensland, St. Lucia, QLD 4072, Australia \\
$^{10}$ School of Physics and Astronomy, Sun Yat-sen University, Zhuhai 519082, China. \\
$^{11}$ American University of Sharjah, Physics Department, P.O. Box 26666, Sharjah, UAE \\
} 
\begin{document} 
\date{Accepted 2019 October 18. Received 2019 October 18; in original form 2019 September 9}
 
\pagerange{\pageref{firstpage}--\pageref{lastpage}} \pubyear{} 
 
\maketitle 
\label{firstpage}

\begin{abstract}
The `chromosome map' diagram (ChM) proved a successful tool to identify and characterize multiple populations
(MPs) in 59 Galactic Globular Clusters (GCs). Here, we construct ChMs for 11 GCs of both Magellanic Clouds (MCs) and with different ages to compare MPs in Galactic and extra-Galactic environments, and explore whether this phenomenon is universal through `place' and `time'.
MPs are detected in five clusters. The fractions of 1G stars, ranging
from $\sim$50\% to $>$80\%, are significantly higher than
those observed in Galactic GCs with similar present-day masses.
By considering both Galactic and MC clusters,
the fraction of 1G stars exhibits: (i) a strong anti-correlation with
the present-day mass, and (ii) with the present-day mass of 2G stars;
(iii) a mild anti-correlation with 1G present-day mass. All Galactic
clusters without MPs have initial masses smaller than $\sim
1.5 \cdot 10^{5}$ M$_{\odot}$ but
a mass threshold governing the occurrence of MPs seems challenged by
massive simple-population MC GCs; (iv) Milky Way clusters with large
perigalactic distances typically host larger fractions of 1G stars,
but the difference disappears when we use initial cluster masses.
These facts are consistent with a scenario where the stars lost by GCs
mostly belong to the 1G.
By exploiting recent work based on Gaia, half of the known Type~II GCs appear
clustered in a distinct region of the integral of motions space, thus
suggesting a common progenitor galaxy.
Except for these Type~II GCs, we do not find any significant
difference in the MPs between clusters associated with different progenitors.
\end{abstract} 
 
 
\section{Introduction}\label{sec:intro}
Observational evidence demonstrates that most Galactic Globular Clusters (GCs) host two main groups of stars with different chemical composition (hereafter first-generation and second-generation, or 1G and 2G). The origin of this phenomenon remains one of the most-intriguing open issues in the field of stellar astrophysics \citep[e.g.][and references therein]{kraft1994a, gratton2004a, gratton2012a, marino2015a, marino2019a}. 

Using the diagnostic tool of a pseudo-color diagram called `Chromosome Map' (ChM), \citet{milone2017a, milone2018a} and \citet{zennarol2019a} have provided an unprecedented detailed analysis of the multiple populations (MPs) pattern in Milky Way GCs.
ChMs have shown that MPs are indeed the common outcome of the formation of all the 59 Milky Way GCs analyzed. 
Although MPs are present in most studied Galactic GCs, each cluster still exhibits its own specific pattern. The number of distinct populations 
 ranges from two (e.g.\,NGC\,6397) to more than seventeen (NGC\,5139, hereafter $\omega$\,Centauri). The
fractions of 1G stars vary from a minimum of less than 10\% ($\omega$\,Centauri) to more than 60\% (e.g.\,NGC\,6717 and NGC\,6838).
Neverthless, MPs in Galactic GCs share some common properties. 
In general,
MPs can be separated into two main discrete groups of 1G and 2G stars in the ChM, and their relative importance depends on the cluster mass. The incidence and complexity of the MP phenomenon in Galactic GCs both increase with cluster mass:
more massive Galactic GCs have larger helium variations and a predominance of 2G stars. Similarly, the color extension of the ChM correlates with the mass of the host GC \citep{milone2015b, milone2017a, milone2018a, lagioia2019b}.
ChMs have also revealed the presence of two classes of clusters, namely Type~I and Type~II GCs, with the latter constituting $\sim$17\% of the objects and displaying a more complex chemical pattern, including variations in Fe and heavy elements \citep[e.g.\,][]{marino2009a, marino2015a, yong2008a, yong2014a, johnson2015a, johnson2017a, dacosta2009a}.
According to some scenarios, GCs have experienced multiple bursts of star formation where 2G stars formed from material polluted by more massive 1G stars \citep[e.g.][]{ventura2001a,decressin2007a,denissenkov2014a,dantona2016a}. \\
One of the most intriguing and controversial implications is that the proto GCs should have been substantially more massive at birth \citep[e.g.][]{ventura2014a}. 
 This condition comes from the evidence that in most clusters the present-day 1G stars are the minority population \citep[e.g.][]{milone2017a} and that only a small fraction of the mass of 1G stars
 is delivered with the proper chemical composition to make 2G stars. Thus, the proto GCs 
should have lost a large fraction of their
1G stars into the Galactic halo, thus making a significant
contribution to the early assembly of the Galaxy \citep[e.g.][]{renzini2015a}. 

In alternative scenarios, all GC stars are coeval and the chemical enrichment of 2G stars is attributed to the accretion of material processed and ejected by massive or supermassive stars of the same generation  in the proto GCs \citep[e.g.][]{bastian2013a,gieles2018a}.

The formation mechanism of the MPs may well have important implications for the assembly of the Milky Way halo and for other galaxies that host GC systems.  Consequently, the present lack of knowledge of the origin of this phenomenon is an important issue that needs to be resolved. What we need to understand is the series of events that led from primordial gas clouds in the early Universe to the GCs with their MPs that we see today; 
if 
the MP phenomenon depends on age (and redshift); and if
any dependence with the parent galaxy exists.
In other words, could the Milky Way environment have favoured the formation of GCs with MPs? And, 
to what extent have GCs contributed to the assembly of their host galaxy, in particular of their stellar halo?

What we know about MPs in 
extragalactic environments comes essentially from observations of GCs in nearby  Local Group galaxies.
Similarly to Galactic GCs, the $\sim 13$-Gyr old GCs of the Large and Small Magellanic Clouds (LMC, SMC), Fornax and the M\,31 GC G\,1 host MPs, \citep[e.g.\,][]{mucciarelli2009a, larsen2014a, hollyhead2017a, niederhofer2017a, lagioia2019a, nardiello2019a, gilligan2019a}, thus indicating that the MP phenomenon is not restricted to the Milky Way.

Spectroscopic elemental abundances of stars in MC clusters younger than $\sim$2 Gyr suggest that these objects, at odds with old Milky Way GCs, are chemically homogeneous \citep[e.g.][]{mucciarelli2014a, martocchia2017a}.
Similarly, the extended main-sequence turn offs and multiple main sequences (MSs) observed in clusters younger than $\sim 2$ Gyr in both the Milky Way and MCs are interpreted as due to stellar rotation\citep[e.g.][]{dantona2015a, milone2017a, li2017a, cordoni2018a, marino2018a, bastian2018a, marino2018b, milone2018b},  rather than to chemical variations, and are possibly associated with age spreads \citep[e.g.][]{goudfrooij2011a, goudfrooij2017a}. These observations have prompted the idea that MPs may have formed  exclusively at high redshift.

On the other hand, the recent discovery   that MC GCs with ages between $\sim$2 and 10 Gyr host stellar populations with different nitrogen abundance \citep[e.g.][]{niederhofer2017a, hollyhead2017a, hollyhead2018a, lagioia2019a} suggests that MPs might not be exotic events from the past, but can also form at lower redshift.
Alternatively, it has been speculated that the occurrence of chemical anomalies may depend on the stellar mass and the MPs appear only in stars with masses less than $\sim$1.6 $\mathcal{M}_{\odot}$ \citep{bastian2018b}, though no specific physical mechanism was envisaged.

However, while many
observations, mostly based on the ChM,
have allowed the accurate characterization of MPs in Galactic GCs, MPs in extragalactic environments are still poorly constrained.
Specifically, the lack of ChMs for extra-Galactic GCs prevents us from a direct comparison
of the MPs properties with Milky Way GCs, and
it remains unclear whether MPs in other galaxies exhibit the same features of variety, discreteness and dependence on the cluster mass as observed in Galactic GCs.

Furthermore, it remains to be understood whether the two classes of Type~I and Type~II GCs have the same origin, or if the Type~II objects could have originated in extragalactic environments, as tentatively suggested by \citet{marino2015a, marino2019a}. 
In this hypothesis, the properties of the parent galaxies hosting Type~II GCs might be different from those of galaxies hosting Type I clusters alone. 
The identification of Type II GCs in the Magellanic Clouds, for example, would be  important to understand the origin of Type II clusters.

To shed light on the dependence of the MP phenomenon on the galactic environment,
an in-depth comparison of young and old GCs in different galaxies
is mandatory to understand to what extent the MPs phenomenon depends on formation redshift and whether the properties of MPs are universal or depend on the host galaxy.
In this paper we exploit multi-band photometry from archive {\it HST} data to extend a similar investigation based on the ChM performed in the surveys of MPs in Galactic GCs to four LMC clusters, namely NGC\,1783, NGC\,1806, NGC\,1846 and NGC\,1978, and seven SMC clusters, Lindsay\,1, Lindsay\,38, Lindsay\,113, NGC\,121, NGC\,339, NGC\,416, and NGC\,419 with ages between $\sim 1.5$ and 10.5 Gyrs.

Historically, several authors \citep[e.g.][]{zinn1993a, vandenberg1993a, mackey2004a} have used metallicities and horizontal-branch morphologies to infer two distinct sub-systems of Milky Way GCs; those formed in satellite galaxies and those formed 'in-situ' within the Galaxy.
In a recent work, \citet{massari2019a}, linked each Galactic GC to the most-probable progenitor galaxy, based on Gaia data release 2 (DR2) data \citep{gaia2018a} and defined two groups of GCs, accreted and  formed in-situ.
To explore the nature of different classes of Milky Way GCs,
we have examined if GCs with different origins 
would exhibit different properties in their stellar populations.

The paper is organized as follows. The data and the data analysis are presented in Section~\ref{sec:data}, while in Sections~\ref{sec:MPs} and~\ref{sec:1G2G} we respectively build the ChMs of the analyzed LMC and SMC clusters and derive the fraction of 1G and 2G stars in each cluster.  Section~\ref{sec:galaxy} is focused on the comparison of stellar populations in GCs of different parent galaxies. Summary and conclusions are provided in Section~\ref{sec:summary}.

\section{Data and data analysis} \label{sec:data} 
In this work we use the fraction of 1G stars with respect to the total number of stars measured for 59 Galactic GCs by \citet{milone2017a, milone2018a} and \citet{zennarol2019a} and analyzed archive data for an additional eleven clusters in the LMC and SMC.

To derive the fraction of 1G stars in the seven Magellanic Cloud clusters Lindsay\,1, Lindsay\,38, Lindsay 113, NGC\,121, NGC\,339, NGC\,416, and NGC\,1978,  we used the photometric and astrometric catalogs published by \citet{lagioia2019b, lagioia2019a} based on {\it HST} images collected through the F336W, F343N, F438W, F814W filters of WFC3/UVIS and the F814W filter of ACS/WFC. We refer to the papers by Lagioia and collaborators for details on the dataset and the data analysis. 

The main properties of the WFC3/UVIS and ACS/WFC images of the other four Magellanic Cloud clusters,  NGC\,419, NGC\,1783, NGC\,1806 and NGC\,1846,  together with additional images of LINDSAY1 and NGC\,416, are summarized in Table~\ref{tab:data}. Stellar photometry and astrometry are derived from images corrected for the poor charge transfer efficiency (CTE) of {\it HST} \citep[see][for details]{anderson2010a} and employing the Jay Anderson's software package KS2, which is the evolution of $kitchen\_sync$, developed by \citet{anderson2008a} to analyze WFC/ACS data. 

Two different methods have been adopted to derive the magnitudes and positions of stars depending on their luminosities. To measure bright stars we first fit the best point-spread function (PSF) model in each individual exposure, and then average the various measurements to get the best estimates for flux and position. 
The KS2 routines combine information from all exposures to measure faint stars, determine the average stellar position from all exposures and then fit each exposure pixel with the PSF solving for the magnitude only \citep[see][for details]{sabbi2016a, bellini2017a}.

Instrumental magnitudes have been calibrated to the Vega mag system as in \citet{bedin2005a} by using the photometric zero points provided by the Space Telescope Science Institute webpages\footnote{\url{http://www.stsci.edu/hst/wfc3/analysis/uvis_zpts/} and \url{http://www.stsci.edu/hst/acs/analysis/zeropoints}}. 
We corrected stellar positions for geometric distortion by using the solutions provided by \citet{anderson2006a} for ACS/WFC or \citet{bellini2009a} and \citet{bellini2011a} for UVIS/WFC3. To select the stars with the best photometry and astrometry, we identified in the catalogs the isolated sources that are well fitted by the PSF model and have small random mean scatters in position and magnitude. To do this, we used the method described by \citet{milone2009a} that is based on the various diagnostics of the astrometric and photometric quality provided by KS2. 
Photometry has been corrected for differential reddening and spatially-dependent variations of the photometric zero point due to small inaccuracies in the sky determination and in the PSF model as in \citet{milone2012a}.

\subsection{Artificial stars}

We further performed artificial-star (AS) experiments to estimate  the photometric errors and to compare the observed ChMs with those expected for a simple stellar population. Specifically, we generated for each cluster a catalog containing positions and fluxes of 100,000 ASs. We assumed for ASs the same F814W luminosity distribution as derived for the real stars and  calculated the corresponding colors from the fiducial lines of red-giant branch (RGB), sub-giant branch (SGB), and MS stars. Moreover, we adopted the same radial distribution for the ASs as observed for the real stars in close analogy to \cite{milone2009a}.  

Photometry and astrometry of ASs have been carried out as in \citet[][see their Section~6]{anderson2008a} by adopting the same computer programs and the same methods by Anderson and collaborators that we used for real stars and described in Section~\ref{sec:data}. 
 We considered an AS as recovered if the measured flux and position differ by less than 0.75 mag and 0.5 pixel, respectively, from the corresponding input values. We applied to ASs the same stringent criteria described in Section~\ref{sec:data} to select a sample of stars with high photometric and astrometric quality and included in the analysis only the selected stars. 
 
\begin{table*}
  \caption{Description of the images of  LINDSAY1, NGC\,416, NGC\,419, NGC\,1783, NGC\,1806, NGC\,1846 from the {\it HST} archive used in the paper.}

\begin{tabular}{c c c c c l l}
\hline \hline
ID & CAMERA &    FILTER  & DATE & N$\times$EXPTIME & PROGRAM & PI \\
\hline
LINDSAY\,1& UVIS/WFC3 &   F275W  & Jun 12 2019 & 1500s$+$1501s$+$2$\times$1523s$+$2$\times$1525s  & 15630 & N.\,Bastian \\  
\hline
NGC\,416  & UVIS/WFC3 &   F275W  & Jun 18 2019 & 1500s$+$1512s$+$2$\times$1525s$+$2$\times$1529s  & 15630 & N.\,Bastian \\  
\hline
NGC\,419  & UVIS/WFC3 &   F336W  & Aug 25 2011 & 400s$+$600s$+$2$\times$700s$+$740s  & 12257 & L.\,Girardi \\  
          & UVIS/WFC3 &   F343N  & Aug 03 2016 & 450s$+$2$\times$1250s$+$1650s       & 14069 & N.\,Bastian \\   
          & UVIS/WFC3 &   F438W  & Aug 03 2016 &  70s$+$150s$+$350s$+$550s           & 14069 & N.\,Bastian \\   
          & WFC/ACS   &   F555W  & Jul 08 2006 & 2$\times$20s$+$4$\times$496s        & 10396 & J.\,Gallagher\\
          & WFC/ACS   &   F814W  & Jan 15 - Jul 08 2006 & 4$\times$10s$+$8$\times$474s  & 10396 & J.\,Gallagher \\
\hline
NGC\,1783 & UVIS/WFC3 &   F336W  & Oct 12 2011 & 2$\times$1190s$+$1200s  & 12257 & L.\,Girardi \\  
          & UVIS/WFC3 &   F343N  & Sep 14 2016 & 450s$+$845s$+$1650s     & 14069 & N.\,Bastian \\   
          & WFC/ACS   &   F435W  & Jan 14 2006 &  90s$+$2$\times$340s    & 10595 & P.\,Goudfrooij \\
          & WFC/ACS   &   F814W  & Jan 14 2006 &   8s$+$2$\times$340s    & 10595 & P.\,Goudfrooij \\
          & WFC/ACS   &   F814W  & Oct 07 2003 & 170s                    &  9891 & G.\,Gilmore \\
    \hline
NGC\,1806    & UVIS/WFC3 & F336W & Oct 12 2011    & 2$\times$1190s$+$1200s& 12275 & L.\,Girardi \\
             & UVIS/WFC3 & F343N & Sep 13-14 2016 & 450s$+$845s$+$1650s   & 14069 & N.\,Bastian \\
             & WFC/ACS   & F435W & Sep 29 2005    & 90s$+$2$\times$340s   & 10595 & P.\,Goudfrooij \\
             & WFC/ACS   & F555W & Sep 29 2005    & 40s$+$2$\times$340s   & 10595 & P.\,Goudfrooij \\
             & WFC/ACS   & F555W & Aug 08 2003    & 300s                  & 9891  & G.\,Gilmore \\
             & WFC/ACS   & F814W & Sep 29 2005    &  8s$+$2$\times$340s   & 10595 & P.\,Goudfrooij \\
             & WFC/ACS   & F814W & Aug 08 2003    & 200s                  & 9891  &  G.\,Gilmore \\
\hline
NGC\,1846    & UVIS/WFC3 & F336W & Apr 16-17 2011 & 900s$+$8$\times$1032s & 12219 & A.\,P.\,Milone \\
             & UVIS/WFC3 & F343N & Apr 04 2016    & 450s$+$845s$+$1650s   & 14069 & N.\,Bastian \\
             & WFC/ACS   & F435W & Jan 01 2006    & 90s$+$2$\times$340s   & 10595 & P.\,Goudfrooij \\
             & WFC/ACS   & F555W & Jan 01 2006    & 40s$+$2$\times$340s   & 10595 & P.\,Goudfrooij \\
             & WFC/ACS   & F555W & Oct 08 2003    & 300s                  & 9891  & G.\,Gilmore \\
             & WFC/ACS   & F814W & Jan 01 2006    &  8s$+$2$\times$340s   & 10595 & P.\,Goudfrooij \\
             & WFC/ACS   & F814W & Oct 08 2003    & 200s                  & 9891  &  G.\,Gilmore \\
     \hline\hline
\end{tabular}
  \label{tab:data}
 \end{table*}

 \subsection{Globular Cluster parameters}\label{sub:gcpar}
 In this paper we investigate stellar populations in Magellanic Cloud and Galactic GCs.
 Our analysis on Galactic GCs requires a number of quantities taken from the literature such as the present-day masses, $\mathcal{M}$, and the initial cluster masses, $\mathcal{M}_{\rm ini}$,  from \citet{baumgardt2018a} and \citet{baumgardt2019a}, the GC ages from \citet{dotter2010a} and \citet{milone2014a}, and the parameters of the GC orbits from \citet{baumgardt2019a}. 
 The integrals of motion (IOM) are provided by \citet{massari2019a} and include the energy, $E$,  the angular momentum in z-direction, $L_{\rm Z}$, and  $L_{\rm PERP}$, which is the angular momentum component perpendicular to $L_{\rm Z}$ \citep[e.g.][]{helmi2000a}. 
 
 We used the fractions of 1G stars derived by \citet{milone2017a} and \citet{milone2018b} and the maximum internal helium variations from \citet{milone2018a} and \citet{zennarol2019a}.  
 We assumed that Ruprecht\,106 and Terzan\,7 are composed of 1G stars alone as shown by \citet{villanova2013a, milone2014a, dotter2018a, lagioia2019b} from either spectroscopy or multi-band photometry of RGB stars. Similarly, we considered AM\,1, Eridanus, Palomar\,3, Palomar\,4, Palomar\,14 and Pyxis as simple populations, as suggested by \citet{milone2014a} based on the horizontal-branch morphology.

The ages and maximum internal helium variations of the eleven LMC and SMC clusters used in this paper are listed in Table~\ref{tab:par} together with the references to the corresponding literature papers.
Present-day masses and initial masses of  NGC\,419, NGC\,1783, NGC\,1806 and NGC\,1846 are taken from  \citet{goudfrooij2014a}, while present-day and initial masses of Lindsay\,1, Lindsay\,38, Lindsay\,113, NGC\,121, NGC\,339, NGC\,416, and NGC\,1978 are derived by Paul Goudfrooij by using the same methods and computer programs described by \citet{goudfrooij2014a} \citep[see also][]{goudfrooij2011a}. 

Although the work by \citet{baumgardt2018a} and \citet{goudfrooij2014a} provide state of the art estimates for initial masses of Galactic and MC GCs, their mass determinations are affected by a number of uncertainties.
An important factor is related to our poor knowledge of the evolution of the Milky Way and
the MCs, and their tidal fields. Other significant uncertainties 
are due to processes during the formation and early evolution of star clusters
whose impacts are hard to estimate quantitatively, especially for star clusters
with current ages older than a few Gyr. Examples of the latter processes are
mass loss due to interactions with molecular clouds in their birth environment
\citep[e.g.,][]{fall2009a,fall2012a} and the unknown level of primordial mass
segregation, which can cause a  significant spread of mass loss rates over the first
few Gyr \citep[e.g.,][]{vesperini2009a}.
 Indeed, GC masses used here are derived by assuming that the gravitational potential of the Galaxy and the MCs is time independent and there is no initial mass segregation. 
In addition, Goudfrooij and collaborators estimate the initial cluster masses of Magellanic Cloud GCs by considering a tidally limited model cluster with a moderate degree of mass segregation, $\mathcal{M}^{\rm seg}_{\rm ini}$, and using the results of the simulation called SG-R1 in \citet{dercole2008a}.
 Table~\ref{tab:par} provides both estimates of GC initial masses.


\section{Multiple populations in Magellanic Cloud clusters}\label{sec:MPs}
The ChM is a pseudo-color diagram used to identify and characterize stellar populations along the MS, RGB, or asymptotic giant branch (AGB) of GCs \citep{milone2015a, marino2017a}. \citet{milone2017a,milone2018a} build the ChMs for 58 GCs by using the $m_{\rm F275W}-m_{\rm F814W}$ color, which is mostly sensitive to stellar populations with different helium abundance, and the $C_{\rm F275W,F336W,F438W}$=($m_{\rm F275W}-m_{\rm F336W}$)$-$($m_{\rm F336W}-m_{\rm F438W}$) pseudo color, which maximizes the separation  among stellar populations with different nitrogen content.
In particular, the ChM allows to distinguish 1G stars, which are distributed around the origin of the reference frame, and 2G stars that are extended towards large $\Delta_{C \rm F275W,F336W,F438W}$ and small (i.e.\,negative) $\Delta_{\rm F275W,F814W}$. 

The fact that accurate photometry in the F275W band can be obtained from space telescopes for relatively bright stars only, is one of the main challenges to derive the ChM of distant GCs.  
To overcame this problem, \citet{zennarol2019a} exploited the $m_{\rm F438W}-m_{\rm F814W}$ color and the $C_{\rm F336W,F343N,F438W}$=($m_{\rm F336W}-m_{\rm F343N}$)$-$($m_{\rm F343N}-m_{\rm F438W}$) pseudo color to build the ChM of the outer-halo  GC NGC\,2419 and of the GCs M\,15 and 47\,Tucanae.  
They find that this alternative ChM, similarly to the classical ChM that involves F275W photometry, is also an efficient tool to identify MPs in GCs and demonstrated that the groups of 1G and 2G stars selected from both ChMs are almost identical. 

To derive the ChM for each analyzed LMC and SMC cluster, we combined information from the $m_{\rm F814W}$ vs.\,$m_{\rm F438W}-m_{\rm F814W}$ CMD and the $m_{\rm F814W}$ vs.\,$C_{\rm F336W,F343N,F438W}$ pseudo-CMD, which are mostly sensitive to stellar populations with different abundances of helium and nitrogen, respectively.  We obtained the verticalized  $\Delta_{\rm F438W,F814W}$ color and the $\Delta_{C {\rm F336W,F343N,F438W}}$ pseudo-color of RGB stars by using the procedure described in \citet{milone2015a, milone2017a} and \citet{zennarol2019a}. 

\begin{centering} 
\begin{figure*} 
  \includegraphics[width=11.0cm,trim={0cm 5cm 5.5cm 3cm},clip]{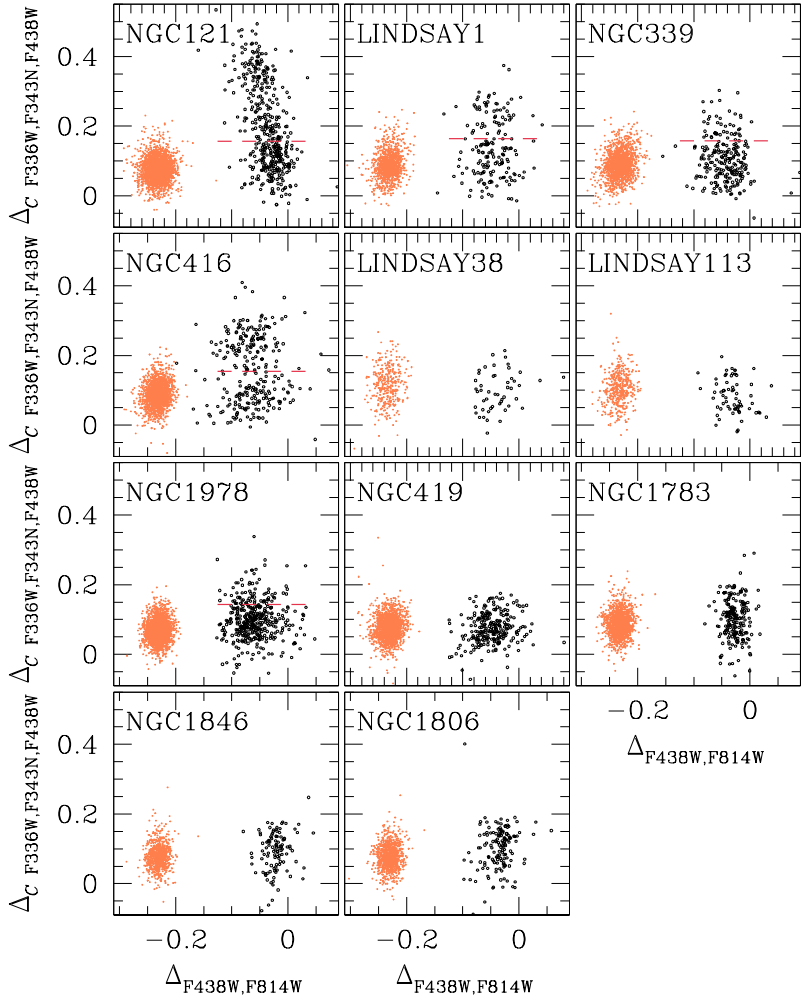}
  \caption{$\Delta_{{\it C} \rm F336W,F343N,F438W}$ vs.\,$\Delta_{\rm F438W,F814W}$ ChMs of the LMC and SMC clusters studied in this paper (black points) and sorted by age from the oldest ($\sim$10.5 Gyr, NGC\,121) to the youngest cluster ($\sim$1.6 Gyr, NGC\,1783). Orange points represent the distribution for a single population and derived from ASs. The scatter is expected from from observational errors alone. The orange points are systematically shifted by $\Delta_{\rm F438W,F814W}$=$-$0.2 mag. The red dashed lines, superimposed on the ChMs of clusters with MPs (NGC\,121, Lindsay\,1, NGC\,339, NGC\,416 and NGC\,1978) correspond to the 95$^{\rm th}$ percentile of the $\Delta_{{\it C} \rm F336W,F343N,F438W}$ distribution of ASs and are indicative of the separation between the bulk of 1G stars and 2G stars.}
 \label{fig:figw} 
\end{figure*} 
\end{centering}

The resulting $\Delta_{C {\rm F336W,F343N,F438W}}$ vs.\,$\Delta_{\rm F438W,F814W}$ ChMs are plotted in Figure~\ref{fig:figw} (black points), where we also show the distribution expected from a single population (orange points) and derived from the ASs. 

Among the eleven analyzed Magellanic Cloud clusters, NGC\,121 exhibits the most-complex ChM. 
In addition to the group of 1G stars clustered around the origin of the reference frame, it hosts an extended 2G that comprises a stellar population with $\Delta_{\rm F438W,F814W} \sim -0.05$ and $\Delta C_{\rm F336W,F343N,F438W} \sim 0.35$ and a group of stars with intermediate values of $\Delta_{\rm F438W,F814W}$ and $C_{\rm F336W,F343N,F438W}$. 
Clearly, the ChMs of Lindsay\,1, NGC\,339 and NGC\,416 exhibit two stellar populations, with the 2G of Lindsay\,1 and NGC\,339 hosting a minority of the total number of cluster stars.  
The distribution of stars in the ChM of NGC\,1978 shows a tail of stars with $\Delta C_{\rm F336W,F343N,F438W} \sim 0.2$, which is not expected from observational errors alone, thus suggesting that this cluster hosts also a small fraction of 2G stars. The remaining GCs, namely Lindsay\,38, Lindsay\,113, NGC\,419, NGC\,1783, NGC\,1806 and NGC\,1846 show no evidence for multiple populations.  The same conclusion that these clusters are consistent with simple populations is provided by \citet{martocchia2019a} by using different photometric diagrams.

\subsection{Reading the $\Delta_{C \rm F336W,F343N,F438W}$ vs.\,$\Delta_{\rm F438W,F814W}$ ChM} 
 To interpret the ChMs shown in the panels of Figure~\ref{fig:figw} we adopted the isochrones used by \citet{milone2018a} that account for the typical chemical composition of stellar populations in GCs. 
  These isochrones are obtained by assuming stellar atmospheric parameters from the Dartmouth stellar evolution database \citep{dotter2008a} and using appropriate synthetic spectra to account for the specific abundances of He, C, N and O in 1G and 2G stars \citep[see][for details]{milone2018a}.  
  
\begin{centering} 
\begin{figure*} 
  \includegraphics[height=7.4cm,trim={0cm 5cm 0cm 10cm},clip]{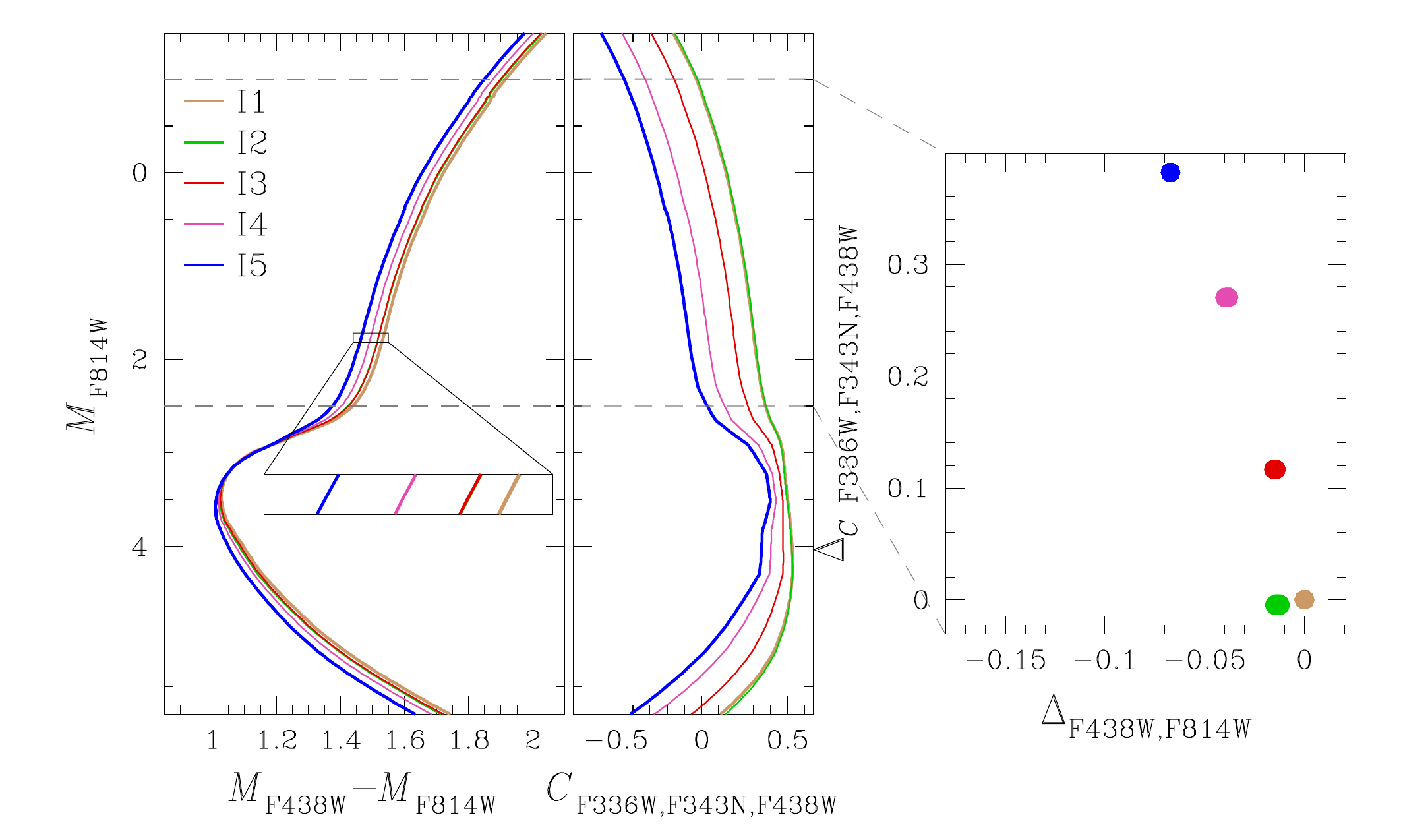}
  \includegraphics[height=7.4cm,trim={12.5cm 5cm 0cm 10cm},clip]{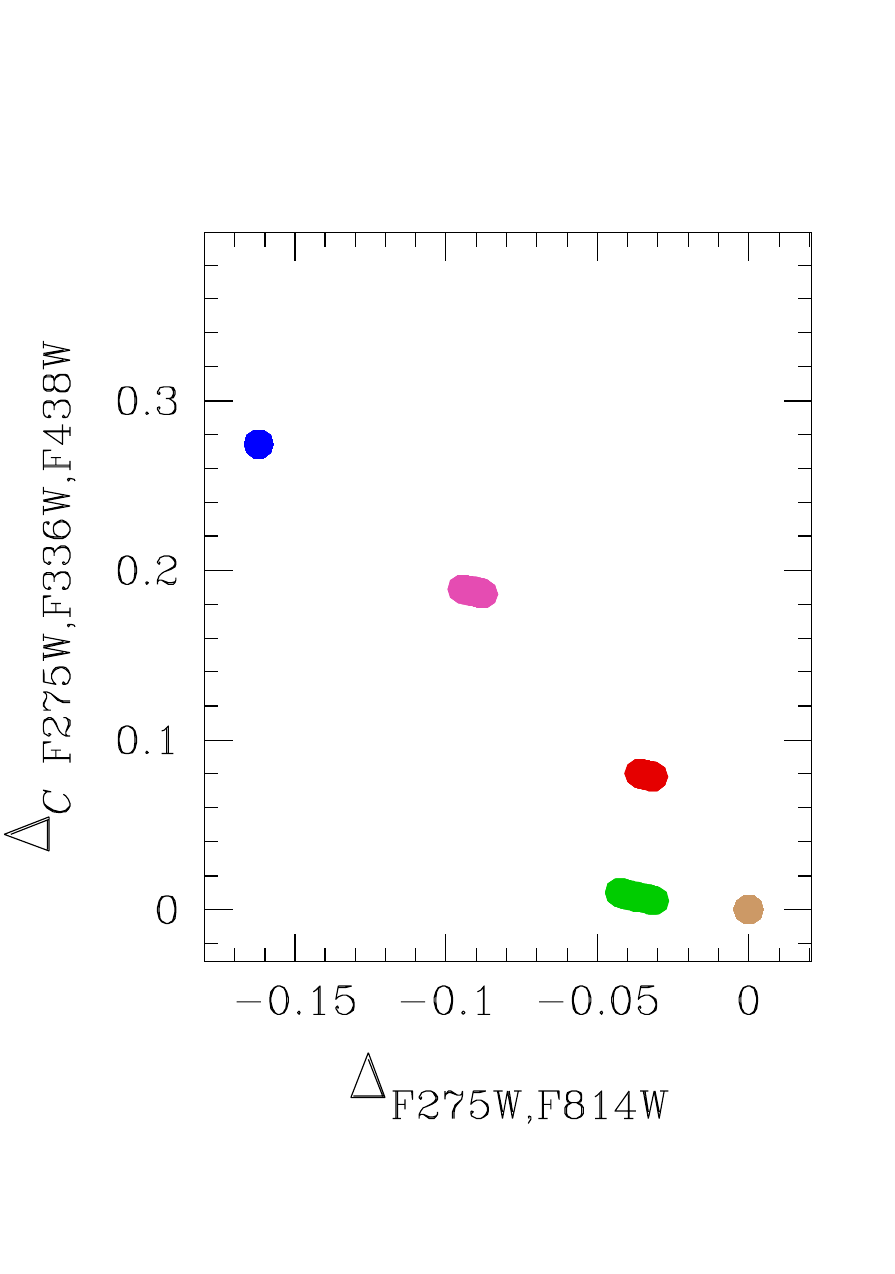}
  \includegraphics[height=7.4cm,trim={0cm 5cm 0cm 10cm},clip]{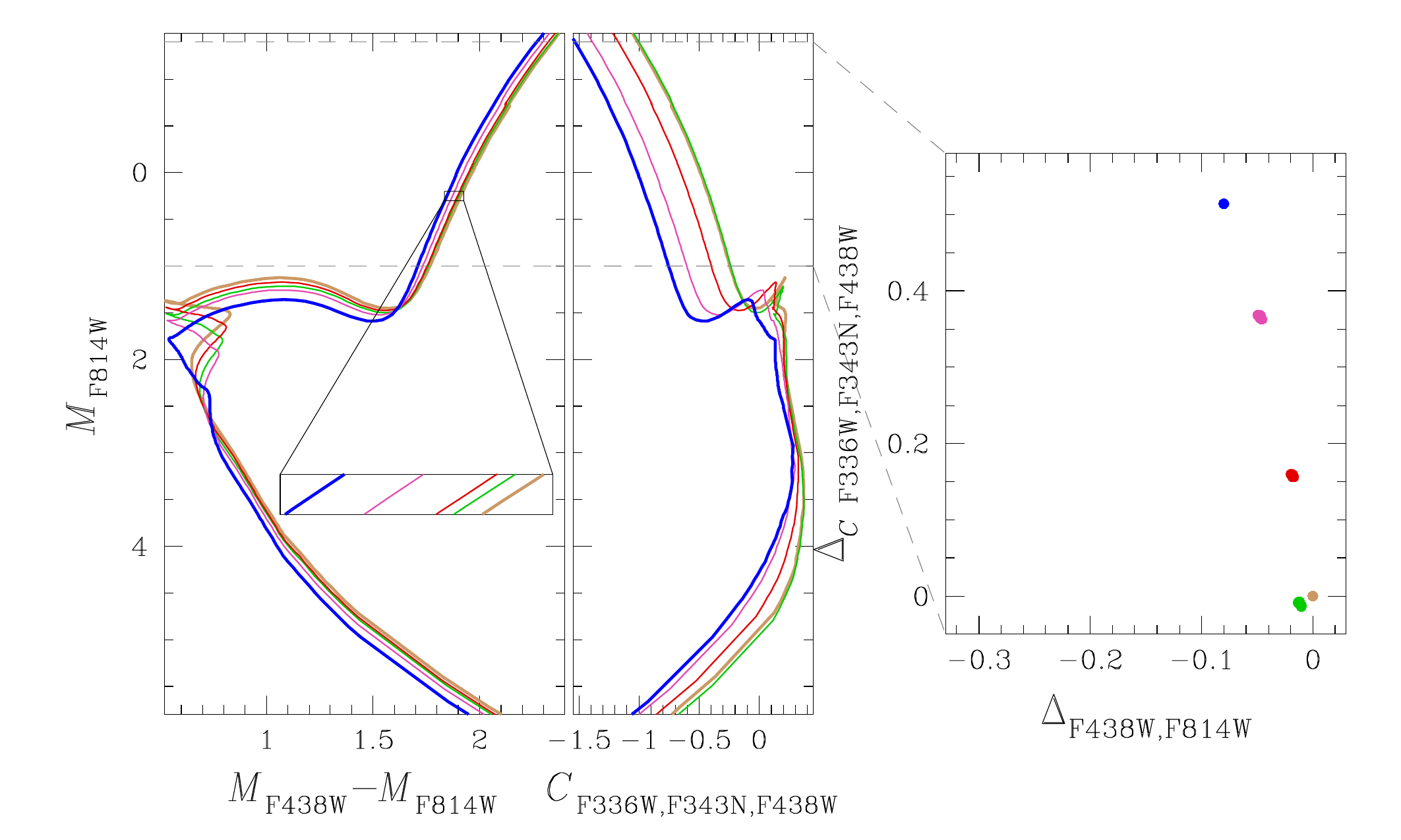}
  \includegraphics[height=7.4cm,trim={12.5cm 5cm 0cm 10cm},clip]{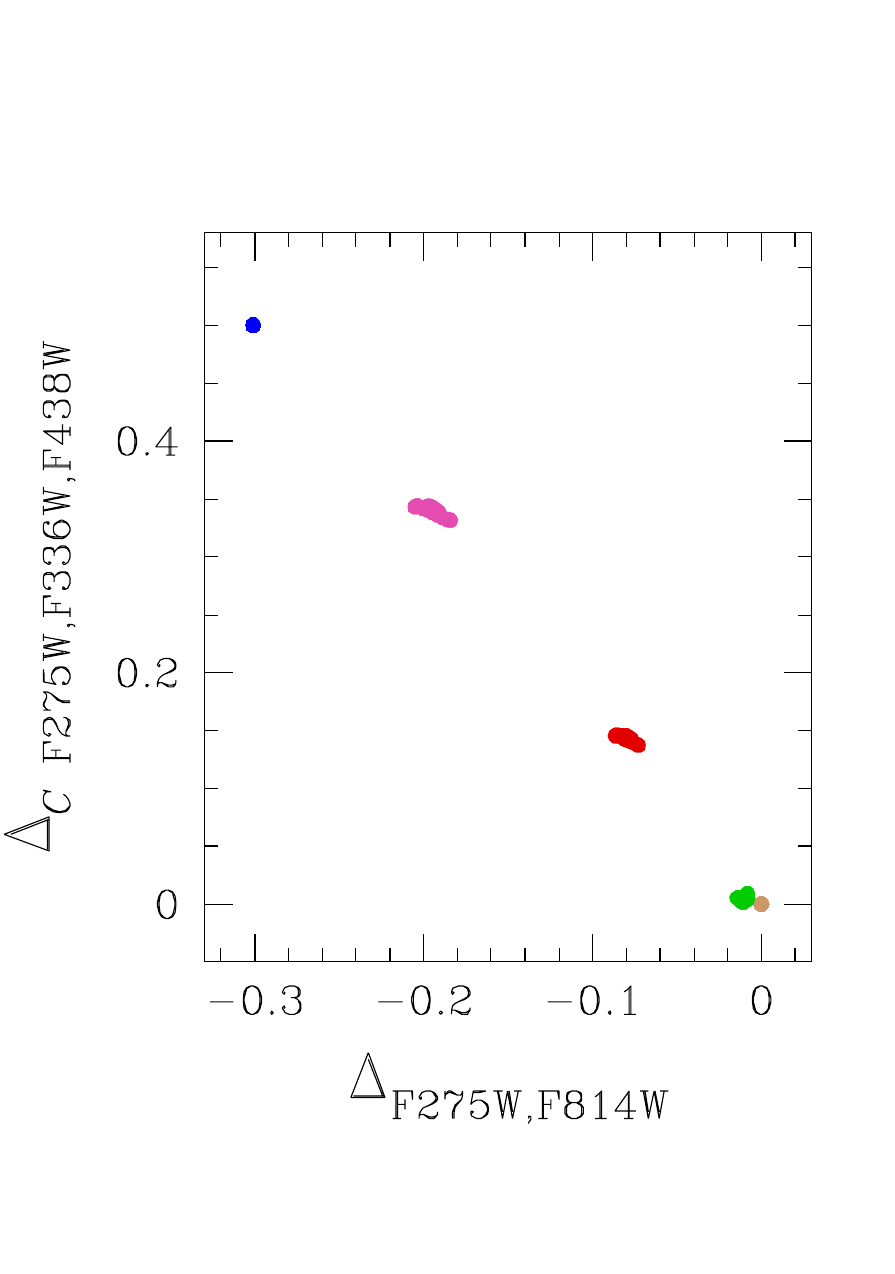}
  \caption{\textit{Upper-Left.} $M_{\rm F814W}$ vs.\,$M_{\rm F438W}-M_{\rm F814W}$ CMD and the $M_{\rm F814W}$ vs.\,$C_{\rm F336W,F343N,F438W}$ pseudo CMD for five 13.0 Gyr old stellar populations with [Fe/H]=$-$1.5, [$\alpha$/Fe]=0.4 and different contents of He, C, N, and O (see text for details).  
  \textit{Upper-Middle.} $\Delta_{{\it C} \rm F336W,F343N,F438W}$ vs.\,$\Delta_{\rm F438W,F814W}$ ChM for stars between the dashed horizontal lines plotted in the left panels. \textit{Upper-Right.}  $\Delta_{{\it C} \rm F275W,F336W,F438W}$ vs.\,$\Delta_{\rm F275W,F814W}$ ChM for the same stars plotted in the middle panel \citep[from][]{milone2018a}. 
  Lower panels show the five 2.0 Gyr old populations with [Fe/H]=$-$0.5, [$\alpha$/Fe]=0.0 and with the corresponding C, N, O abundances.
   Note that, while I1 and I5 isochrones translate into a single point in the ChMs, the remaining populations span a small but significant range of  $\Delta_{{\it C} \rm F336W,F343N,F438W}$, $\Delta_{\rm F438W,F814W}$, $\Delta_{{\it C} \rm F275W,F336W,F438W}$  and $\Delta_{\rm F275W,F814W}$ in the corresponding ChMs.  
   }
 \label{fig:ChMz} 
\end{figure*} 
\end{centering}

 We plot in the left panels of Figure~\ref{fig:ChMz} the $M_{\rm F814W}$ vs.\,$M_{\rm F438W}-M_{\rm F814W}$ CMD and the $M_{\rm F814W}$ vs.\,$C_{\rm F336W,F343N,F438W}$ pseudo CMD of five isochrones, I1---I5, with ages of 13.0 Gyr, [Fe/H]=$-$1.5, [$\alpha$/Fe]=0.4 and different chemical compositions. 
 Specifically, we assumed that the two isochrones I1 and I5 have extreme helium values of Y=0.247 and 0.297, while isochrones I2, I3 and I4 have intermediate helium abundances and are enhanced in helium by ∆Y=0.02, 0.02, and 0.03, respectively, with respect to I1.
  We used solar C, N and O abundances for both I1 and I2. I5 has the most-extreme chemical composition
   ([C/Fe]=$-$0.5, [N/Fe]=1.2 and [O/Fe]=$-0.1$), while I3 and I4 exhibit intermediate C, N, O  abundances that correspond to [C/Fe]=$-$0.15, [N/Fe]=0.40 and [O/Fe]=0.20 and [C/Fe]=$-$0.30, [N/Fe]=0.80 and [O/Fe]=$0.05$, respectively.
 
  The isochrone segments with $-1.0 <M_{\rm F814W}< 2.5$ are used to derive the $\Delta_{{\it C} \rm F336W,F343N,F438W}$ vs.\,$\Delta_{\rm F438W,F814W}$ ChM plotted in the middle panels of Figure~\ref{fig:ChMz}. 
  The isochrone I1 populates the origin of the reference frame in this ChM, whereas stars with the most-extreme chemical composition (isochrone I5) exhibit the largest values of    $\Delta_{\rm F438W,F814W} \sim -0.07$ and $\Delta_{{\it C} \rm F336W,F343N,F438W} \sim 0.39$. The stellar populations described by isochrones I3 and I4 occupy intermediate positions in the ChMs with respect to I1 and I5. Isochrones I1 and I2, which share the same C, N, and O abundances but different helium content, have similar $\Delta_{{\it C} \rm F336W,F343N,F438W}$ values.    \\
  The isochrones plotted in Figure~\ref{fig:ChMz} are previously used by \citet{milone2018a} to characterize the classical $\Delta_{{\it C} \rm F275W,F336W,F438W}$ vs.\,$\Delta_{\rm F275W,F814W}$ ChM (see their Figure~6). We reproduce the ChM by Milone and collaborators in the right panels of Figure~\ref{fig:ChMz}. 
   
  The $\Delta_{\rm F275W,F814W}$ quantity used in the classical ChM provides a wider pseudo-color separation between the stellar populations than $\Delta_{\rm F438W,F814W}$. This is mostly due to the fact that stars with the same luminosity but different helium contents differ in their effective temperatures. As a consequence, the $M_{\rm F275W}-M_{\rm F814W}$ color, which is more sensitive than $M_{\rm F438W}-M_{\rm F814W}$ to effective-temperature variations, provides a wider pseudo-color separation among stellar populations with different helium abundance. 
  On the other hand, the $\Delta_{{\it C} \rm F275W,F336W,F438W}$ pseudo-color is slightly less effective than $\Delta_{{\it C} \rm F336W,F343N,F438W}$ to separate the five stellar populations plotted in Figure~\ref{fig:ChMz}. 
  
    Despite the differences above, the comparison between middle and left panels in Figure~\ref{fig:ChMz} reveals that, qualitatively, the five stellar populations occupy similar relative positions in both ChMs. This fact confirms previous findings by \citet{zennarol2019a} who show that 1G and 2G stars in NGC\,104 and NGC\,7078 occupy similar positions in the $\Delta_{{\it C} \rm F336W,F343N,F438W}$ vs.\,$\Delta_{\rm F438W,F814W}$ and $\Delta_{{\it C} \rm F275W,F336W,F438W}$ vs.\,$\Delta_{\rm F275W,F814W}$ ChMs. 

In the lower panels of Figure~\ref{fig:ChMz} we extend the analysis to five stellar populations with [Fe/H]=$-$0.5, [$\alpha$/Fe]=0.0 and age of 2~Gyr. The differences in He, C, N and O for the isochrones I2--I5 with respect to I1 are the same as those adopted for the simulation of the 13-Gyr stellar populations.  
A visual inspection of the classical ChM and the $\Delta_{{\it C} \rm F336W,F343N,F438W}$  vs.\,$\Delta_{\rm F438W,F814W}$ ChM reveals that MPs in young and old GCs qualitatively exhibit similar behaviours. We also note that 2-Gyr old stellar populations span a wider range of  $\Delta_{{\it C} \rm F336W,F343N,F438W}$ and $\Delta_{\rm F438W,F814W}$ than those with 13 Gyr. Such difference is mostly due to the fact that the young populations are more-metal rich than the old ones.  Moreover, the RGBs of the I2 and I3 stellar populations plotted in the upper panel of Figure~\ref{fig:ChMz} share almost the same F438W$-$F814W colors, while the I3 RGB stars shown in the lower panel have slightly bluer F438W$-$F814W colors than I2 stars with the same F814W magnitude. We verified that the different metallicities and effective temperatures of the analyzed young and old RGB stars are responsible for such small color difference.

\begin{centering} 
\begin{figure*} 
  \includegraphics[width=13cm,trim={0cm 6.5cm 0cm 15.1cm},clip]{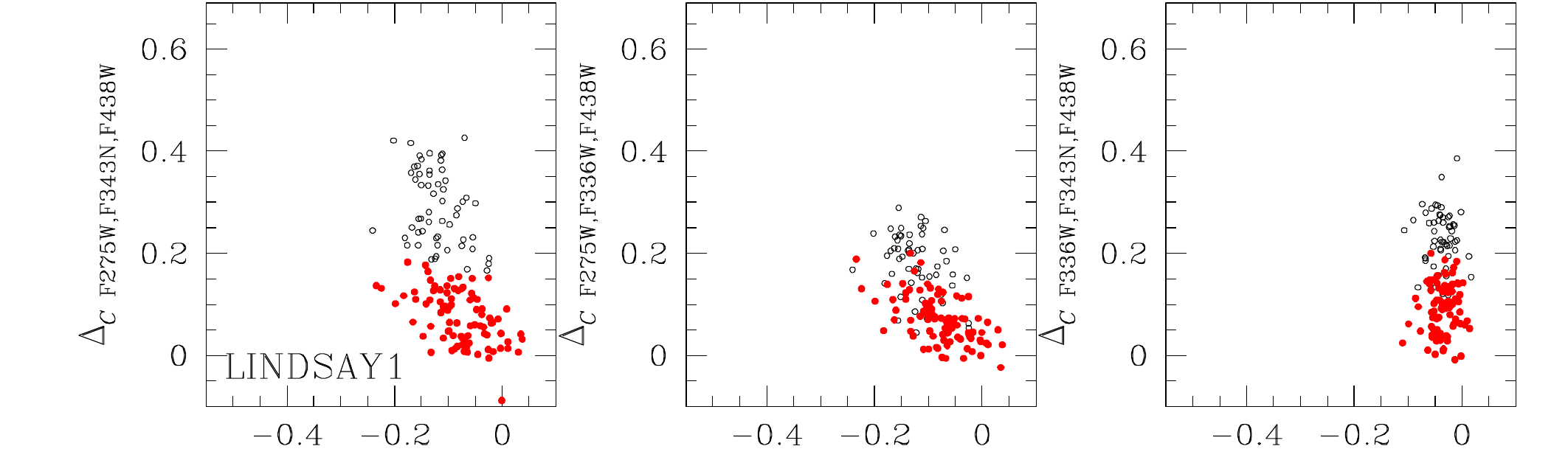}
  \includegraphics[width=13cm,trim={0cm 5cm 0cm 15.1cm},clip]{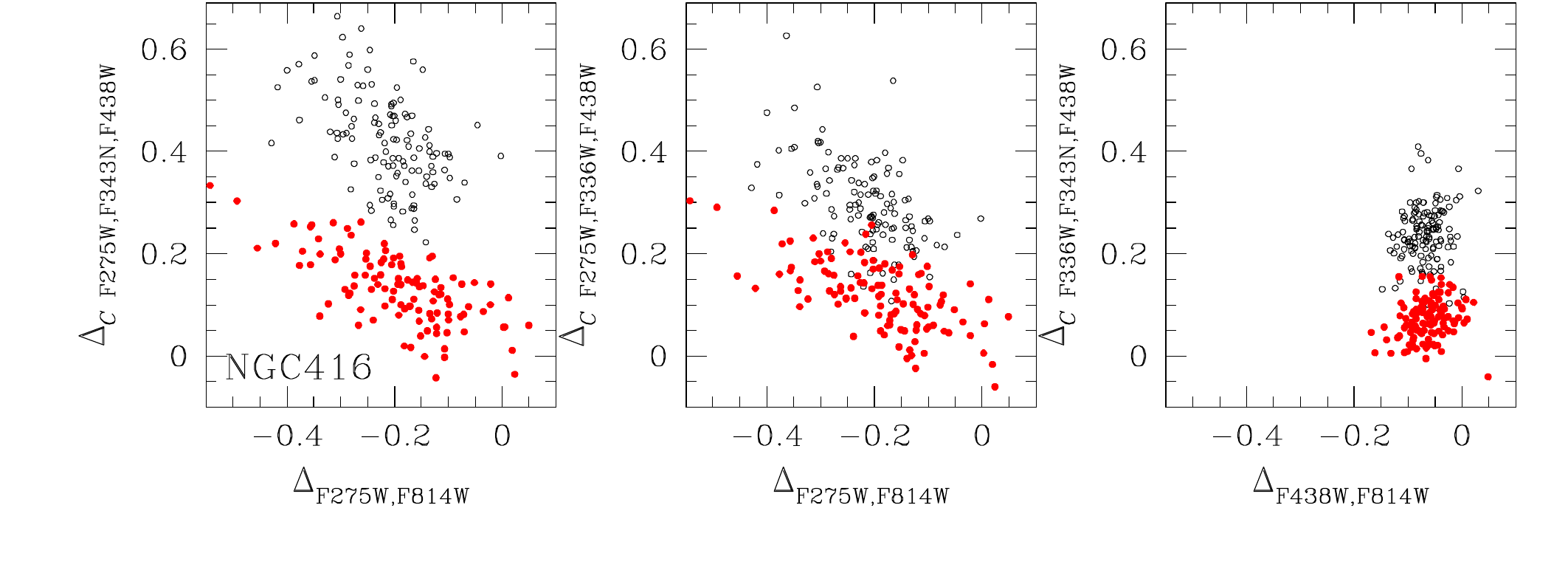}
  \caption{ Comparison of the $\Delta_{{\it C} \rm F275W,F343N,F438W}$ vs.\,$\Delta_{\rm F275W,F814W}$ (left), $\Delta_{{\it C} \rm F275W,F336W,F438W}$ vs.\,$\Delta_{\rm F275W,F814W}$ (middle) and $\Delta_{{\it C} \rm F336W,F343N,F438W}$ vs.\,$\Delta_{\rm F438W,F814W}$ ChMs (right) for Lindsay\,1 and NGC\,416. Red and black colors indicate bona-fide 1G and 2G stars, respectively, selected from the left-panel ChMs.}
 \label{fig:ChM275} 
\end{figure*} 
\end{centering}

 To further demonstrate that the 1G and 2G stars occupy similar positions in the ChMs by \citet{milone2017a} and in those used in this paper, we compare in Figure~\ref{fig:ChM275} the $\Delta_{{\it C} \rm F275W,F343N,F438W}$ vs.\,$\Delta_{\rm F275W,F814W}$ \citep[see][]{milone2019a}, $\Delta_{{\it C} \rm F275W,F336W,F438W}$ vs.\,$\Delta_{\rm F275W,F814W}$ and $\Delta_{{\it C} \rm F336W,F343N,F438W}$ vs.\,$\Delta_{\rm F438W,F814W}$ ChMs for Lindsay\,1 and NGC\,416. Clearly, the sample of 1G stars and 2G stars, which are identified from left-panel ChMs and are colored red and black, respectively, exhibit similar relative locations in the three ChMs.
  These results demonstrate that the $\Delta_{{\it C} \rm F336W,F343N,F438W}$ vs.\,$\Delta_{\rm F438W,F814W}$ diagram is an efficient tool to identify and characterize 1G and 2G stars in young and old GCs.

\begin{centering} 
\begin{figure*} 
  \includegraphics[height=7cm,trim={0cm 5cm 0cm 10cm},clip]{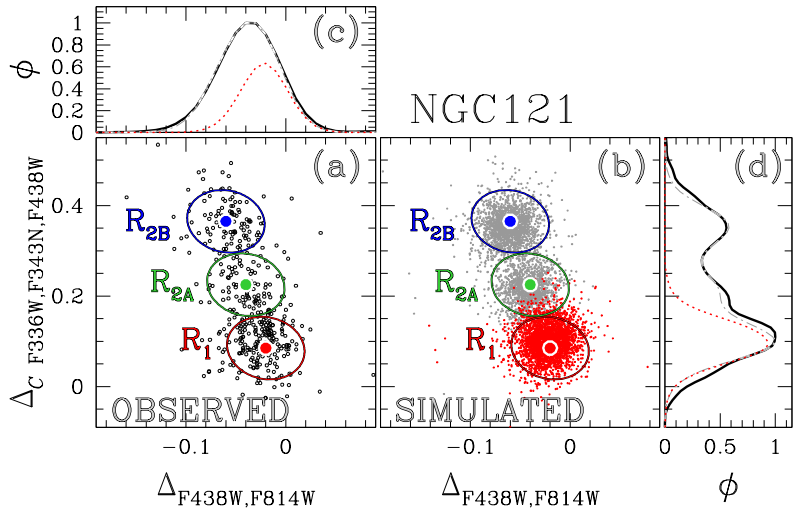}
  \caption{Procedure to estimate the fraction of 1G, 2G$_{\rm A}$ and 2G$_{\rm B}$ stars in NGC\,121. The observed ChM is reproduced in panel a, while panel b shows the simulated ChM where simulated 1G stars are colored in red. The three elliptical regions, R$_{1}$,  R$_{\rm 2A}$ and  R$_{\rm 2B}$ that we used to estimate the population ratios are superimposed on both ChMs and are colored in red, green and blue, respectively. Continuous black lines and gray dashed-dotted lines plotted in panels c and d show the $\Delta_{\rm F438W,F814W}$ and $\Delta_{ C \rm F336W,F343N,F438W}$ kernel-density distributions of stars in the observed and simulated ChM, respectively. Red-dotted lines are the kernel-density distributions of simulated 1G stars.}
 \label{fig:n121pop} 
\end{figure*} 
\end{centering} 

\begin{centering} 
\begin{figure*} 
  \includegraphics[height=7cm,trim={0cm 5cm 0cm 10cm},clip]{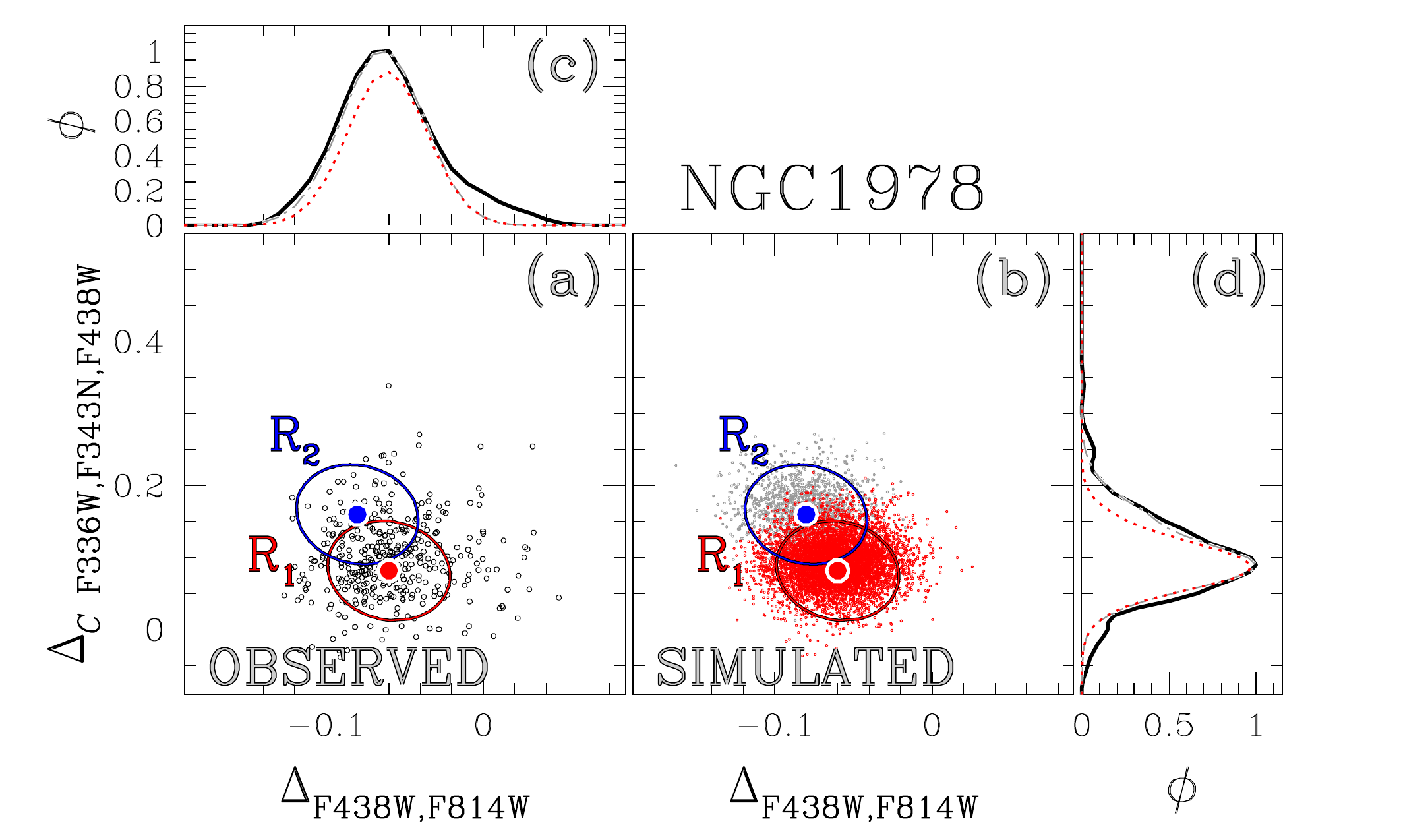}
  \caption{As in Figure~\ref{fig:n121pop} but for NGC\,1978, which is the cluster where the separation between 1G and 2G stars is less evident.}
 \label{fig:n1978pop} 
\end{figure*} 
\end{centering} 

\section{The fraction of first- and second-generation stars in LMC and SMC clusters}\label{sec:1G2G}
To calculate the relative numbers of stars in the stellar populations of  Lindsay\,1, NGC\,121, NGC\,416, NGC\,339, and NGC\,1978 we applied the procedure illustrated in Figure~\ref{fig:n121pop} for NGC\,121. This method was developed by \citet{milone2012b} to derive the fraction of stars in the distinct stellar populations in GCs by using the CMD and extended by \citet{nardiello2018a} and \citet{zennarol2019a} to the ChM.  

We defined three elliptical regions, namely $R_{1}$, $R_{2}$, and $R_{3}$ in the ChM of NGC\,121. Each ellipse has the same axis ratio and inclination as the ellipse that provides the best fit with the error distribution.  The coordinates of the centers of the ellipses correspond to the average values of $\Delta_{\rm F438W,F814W}$ and $\Delta_{C {\rm F336W,F343N,F438W}}$ of the stars in the three stellar populations of the clusters. %

We expect that, due to observational errors, each region comprises stars of the three stellar populations. Specifically, the number of stars within the region $R_{1}$, $N_{1}$, is made up of 1G, 2G$_{\rm A}$ and 2G$_{\rm B}$ stars ($N_{\rm 1G}$, $N_{\rm 2GA}$, $N_{\rm 2GB}$) according to the following relation
\begin{equation}\label{eq:1}
 N_{1}=N_{\rm 1G} f_{\rm 1G}^{\rm R1}    +  N_{\rm 2GA} f_{\rm 2GA}^{\rm R1} + N_{\rm 2GB} f_{\rm 2GB}^{\rm R1} 
 \end{equation}
where  $f_{\rm 1G}^{\rm R1}$, $f_{\rm 2GA}^{\rm R1}$, and $f_{\rm 2GB}^{\rm R1}$  are the fractions of 1G, 2G$_{\rm A}$ and 2G$_{\rm B}$ stars, respectively, within R1.
 Similar relations involve the number of stars, $N_{2}$ ($N_{3}$), within the regions $R_{2}$ ($R_{3}$) of the ChM and the fraction of 1G, 2G$_{\rm A}$ and 2G$_{\rm B}$ stars within these regions ($f_{\rm 1G}^{\rm R2 (R3)}$, $f_{\rm 2GA}^{\rm R2 (R3)}$, and $f_{\rm 2GB}^{\rm R2 (R3)}$).
 
 The number of stars in the three populations of NGC\,121 are calculated by solving for these three equations. 
 The values of $N_{1}$, $N_{2}$ and $N_{3}$ used in these equations are derived by counting the stars within the corresponding ellipses, as illustrated in Figure~\ref{fig:n121pop}a. In Figure~\ref{fig:n121pop}b we show the procedure to derive the fraction of 1G stars within the R1, R2 and R3 used in the equations~\ref{eq:1}.  To do this, we used ASs to simulate 1G stars only (red points). The values of $f_{\rm 1G}^{\rm R1}$, $f_{\rm 2GA}^{\rm R1}$, and $f_{\rm 2GB}^{\rm R1}$ are the ratio between the number of simulated stars within R1, R2 and R3 and the total number of simulated 1G stars. We used a similar procedure to estimate the fractions of 1G, 2G$_{\rm A}$ and 2G$_{\rm B}$ stars within regions R$_{\rm 2A}$ and R$_{\rm 2B}$.  

 We find that 1G includes 51.7$\pm$\%2.6 of the total number of analyzed RGB stars, while the 2G$_{\rm A}$ and 2G$_{\rm B}$ host 20.7$\pm$2.1\% and 27.6$\pm$2.3\% of stars, respectively. The size of each ellipse is chosen by eye with the criterium of including at the same time the bulk of simulated ASs and minimizing the overlap between two adjacent regions. However, we verified that by changing the major axis by $\pm$25\% the resulting values of the population ratios are the same within 0.008 \citep[see also][]{nardiello2018a}. Similarly, we verified that when we use circles instead of ellipses the results are nearly the same, thus indicating that the conclusion do not significantly depend on the shape of the region.
 
The same method has been applied to Lindsay\,1, NGC\,416, NGC\,339 and NGC\,1978, where we identified two groups of 1G and 2G stars. 
Figure~\ref{fig:n1978pop} illustrates the same procedure described above for NGC\,121 but for NGC\,1978, which is the MP cluster where the 2G is less evident.
 Results are listed in Table~\ref{tab:par}. The fraction of 2G stars dramatically changes from one cluster to another and ranges from $\sim$0.15 in NGC\,1978 and NGC\,339 to $\sim$0.50 in NGC\,121.

\subsection{Internal helium variation}
 We verified that the groups of 1G and 2G stars of Lindsay\,1, NGC\,121, NGC\,339 and NGC\,416 identified in our paper comprise almost the same stars classified as first- and second generation by \citet{lagioia2019a} who also estimated the helium difference between 2G and 1G stars in  these GCs ($\Delta$Y$_{\rm 2G-1G}$).  Since Lindsay\,1, NGC\,339 and NGC\,416 host only two populations, the values of $\Delta$Y$_{\rm 2G-1G}$ derived by Lagioia and collaborators correspond to the maximum internal helium variation within these clusters ($\Delta$Y$_{\rm max}$). 

In the case of NGC\,121, we estimated $\Delta$Y$_{\rm max}$ as the helium difference between 2G$_{\rm B}$ and 1G stars by using the method introduced in \citet{milone2013a} and used by \citet{lagioia2019a}. Briefly, we measured the $m_{\rm X}-m_{\rm F814W}$ color difference between the RGB fiducial lines of 2G$_{\rm B}$ and 1G stars at four different values of the F814W luminosity. The comparison between the observed colors with appropriate grid of synthetic spectra with different abundances of He, C, N and O provides an estimate of the relative abundances of these elements. We find that the maximum internal helium variation within NGC\,121 is $\Delta$Y$_{\rm max}=0.014 \pm 0.004$, and the abundance differences between the synthetic spectra of 2G$_{\rm B}$ and 1G stars that provide the best fit with the data are  $\Delta$[C/Fe]=$-$0.30$\pm$0.10, $\Delta$[N/Fe]=0.80$\pm$0.15 and $\Delta$[O/Fe]=$-$0.45$\pm$0.15.  Similarly, we estimate that the helium difference between the two stellar populations of NGC\,1978 is $\Delta$ Y$_{\rm max}=0.002 \pm 0.003$, and the best-fit values of $\Delta$[C/Fe], $\Delta$[N/Fe] and $\Delta$[O/Fe] correspond to $-$0.05$\pm$0.05, 0.07$\pm$0.03 and 0.00$\pm$0.03, respectively. 

In the same way as we have distinguished between the classical phenomenon of MPs in ancient GCs and the presence of extended turnoffs and split main sequences of young MC and Galactic clusters (attributed to stellar populations with different rotation rates and possibly different ages) a note of warning is necessary when discussing the nitrogen difference among stars. In fact, a nitrogen increase as low as found among the stars in NGC\,1978 is not necessarily a signature of `multiple populations' in the sense of the term derived from our knowledge of ancient GCs. For instance, in the context of binary evolution, pure CN cycling, occurred in the stellar interior and later exposed in the stellar atmosphere of a companion star through different paths of binary evolution, can account for up to a factor two increase in the nitrogen abundance.  Further, the small percentage of these anomalous stars ($\sim$15\%) makes them compatible with being remnants of peculiar evolutionary paths. Binary evolution, on the contrary can not explain [N/Fe] variations larger than $\sim$0.3 dex associated with oxygen depletion, which are associated with the full CNO cycle.  

\section{Relations with the host galaxy}\label{sec:galaxy}
To investigate the importance of environment and age on the onset of MPs in GCs and to understand whether the MP properties depend on the host galaxy or not, we compare the results on Magellanic Cloud clusters established in the previous section, with similar findings for Galactic GCs. 

\subsection{Comparing multiple populations of Galactic and Magellanic-Cloud clusters}
As discussed in Section~\ref{sub:gcpar}, work based on the ChM provided homogeneous determinations of the fractions of 1G and 2G stars in 59 clusters \citep{milone2017a, zennarol2019a} mostly observed as part the UV survey of Galactic GCs \citep{piotto2015a}. 
Besides these 59 GCs, some clusters seem to show no evidence of multiple populations \citep[see][for details]{villanova2013a, milone2014a, dotter2018a, lagioia2019b}. 

\begin{centering} 
\begin{figure} 
  \includegraphics[height=8.2cm,trim={.5cm 5cm 0cm 4.5cm},clip]{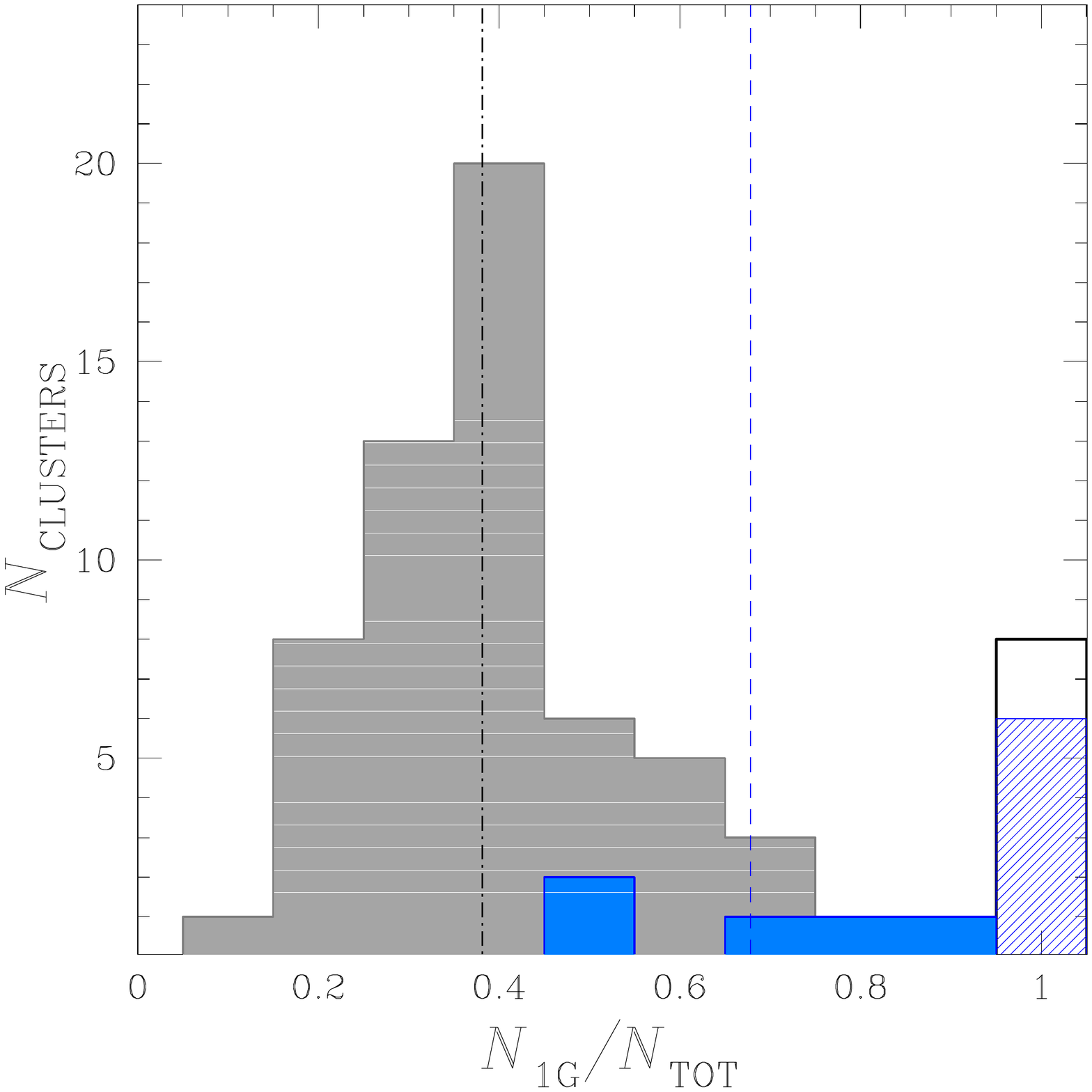}
  \caption{Histogram distributions of the fraction of 1G stars in Galactic GCs with multiple populations (gray histogram) and without multiple populations (black open histogram). Multiple-population and single-population Magellanic Cloud clusters are represented with blue histograms and white-blue histograms, respectively. The gray dashed-dotted line at $N_{\rm 1G}/N_{\rm TOT}=0.38$ and the blue dashed line at $N_{\rm 1G}/N_{\rm TOT}=0.68$ marks the average fractions of 1G stars for Galactic and Magellanic Cloud clusters with multiple populations, respectively.}
 \label{fig:histo} 
\end{figure} 
\end{centering}

We compare in Figure~\ref{fig:histo} the histogram distribution of the fraction of 1G stars in Galactic GCs (gray histogram), with the corresponding distribution for SMC and LMC GCs (blue histogram). The analyzed LMC and SMC clusters with multiple populations host, on average, a higher fraction of 1G stars than the studied sample of Galactic GCs. In particular, the fractions of 1G stars of NGC\,339 and NGC\,1978 are significantly larger than those observed in Milky Way GCs.

\citet{milone2017a} showed that the fractions of 2G stars in Galactic GCs correlate with the absolute luminosity of the host cluster \citep[from the 2010 version of the][catalog]{harris1996a} and with cluster mass \citep[from][]{mclaughlin2005a}. Although there is no significant correlation with the orbital parameters of the host clusters, GCs with large perigalactic distances ($R_{\rm PER}>3.5$ kpc) tend to have smaller fractions of 2G stars than GCs with similar absolute luminosities and $R_{\rm PER}<3.5$ kpc  \citep{zennarol2019a}. We verified that the result by \citet{zennarol2019a}, which is based on the values of $R_{\rm PER}$ by \citet{baumgardt2018a}, is confirmed when perigalactic distances from \citet{massari2019a} are used.

\begin{centering} 
\begin{figure*} 
  \includegraphics[width=10.cm,trim={.4cm 5.8cm 0cm 6.0cm},clip]{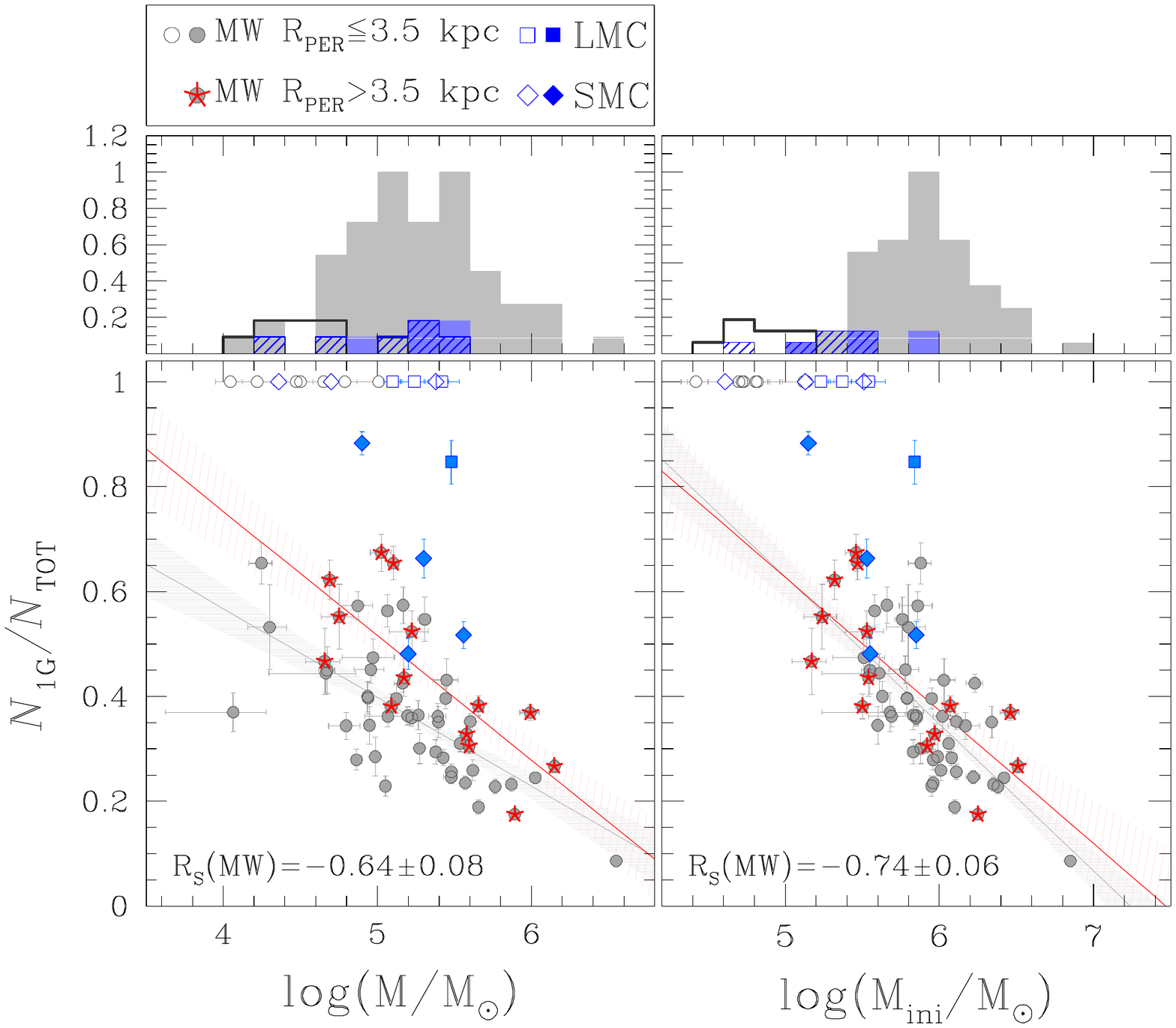}
  \includegraphics[width=10.cm,trim={.4cm 5.8cm 0cm 12.4cm},clip]{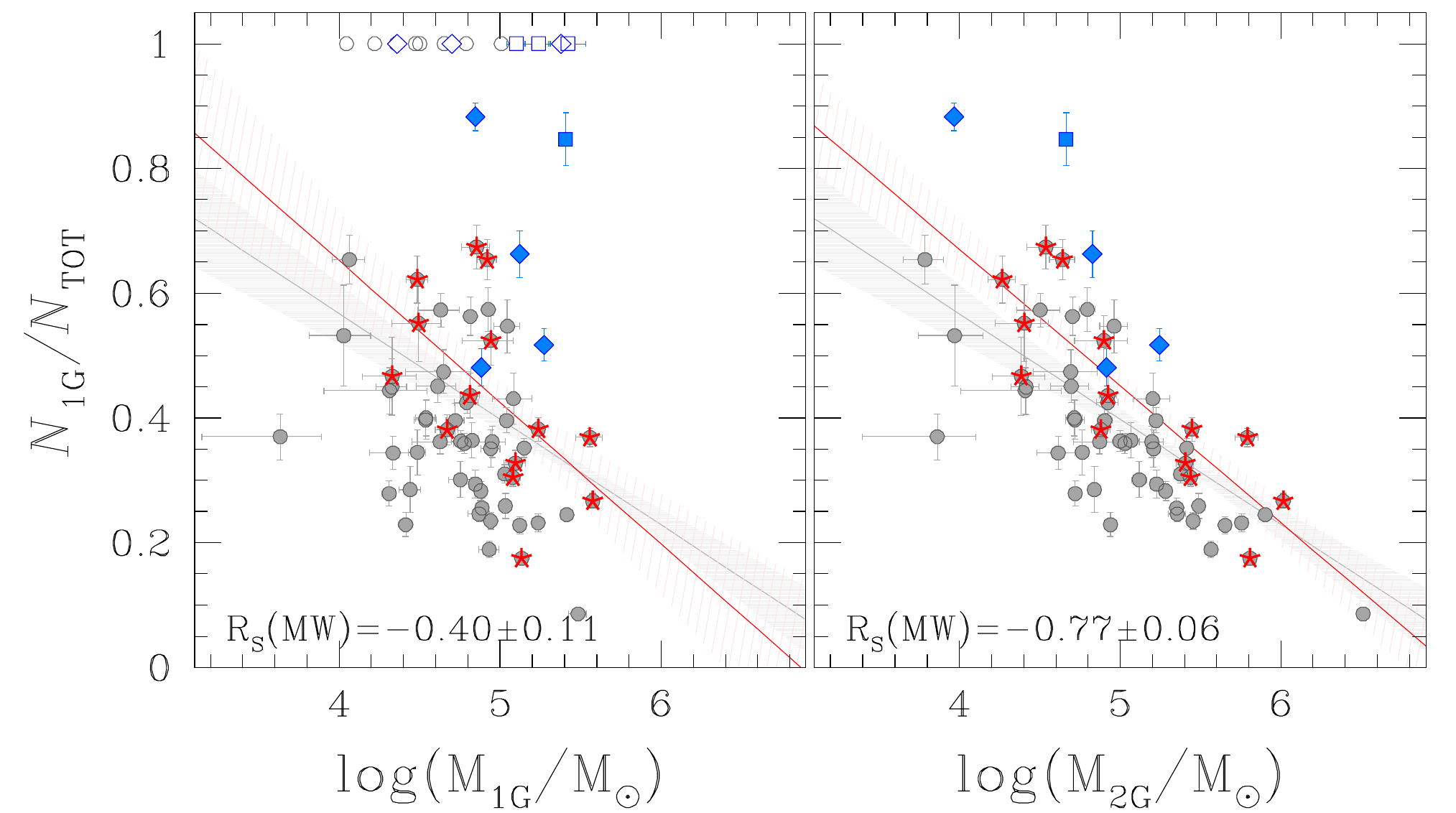}
  \caption{Fraction of 1G stars against the present-day cluster mass (upper-left) and the initial cluster mass (upper-right) for Galactic GCs (gray dots), LMC (blue squares) and SMC (blue diamonds) clusters. Open symbols with  $N_{\rm 1G}/N_{\rm TOT}$=1.0  indicate GCs with no evidence of multiple populations while Galactic GCs with perigalactic radius, $R_{\rm PER}>3.5$ kpc are marked with red starred symbols. The histogram distributions of $\log{\mathcal{M}/\mathcal{M_{\odot}}}$ and $\log{\mathcal{M_{\rm ini}}/\mathcal{M_{\odot}}}$ are also shown. Gray and blue shaded histograms indicate Milky Way GCs and Magellanic Cloud GCs with detected MPs, respectively. Black open histograms and white-blue histograms correspond to Galactic and Magellanic Cloud clusters with no evidence of multiple populations.
   Lower panels show the fraction of 1G stars against the present-day mass of the 1G (left) and the 2G (right). For illustration purposes simple-population GCs, where the mass of the 2G is equal to zero, are not plotted in the bottom-right panel.
The least-square best fit lines derived from Galactic GCs with  $R_{\rm PER}\lesssim3.5$ kpc and $R_{\rm PER}>3.5$ kpc are colored gray and red, respectively, while the corresponding shaded areas include the 68.27\% confidence interval of the true regression line. We quoted in each panel the Spearman's rank correlation coefficient calculated for Milky Way GCs alone, R$_{\rm S}$(MW).     
    }
 \label{fig:2Gmass} 
\end{figure*} 
\end{centering}

In the upper-left panel of Figure\,\ref{fig:2Gmass} we adopted gray dots to represent the fraction of 1G stars for Galactic GCs against the logarithm of the GC masses by \citet{baumgardt2018a}. When we consider only Galactic GCs with multiple populations, the Spearman's rank correlation coefficient between these two quantities is $R_{\rm s}=-0.64 \pm 0.08$, where the uncertainty is estimated by bootstrapping with replacements performed 10,000 times. The error indicates one standard deviation of the bootstrapped measurements. 
Hence, we confirm the anti-correlation between GC mass and  fraction of 1G stars for clusters with MPs.
The Spearman's rank correlation coefficient is $R_{\rm s}=-0.72 \pm 0.06$ when we include the analyzed candidate simple population GCs.

We confirm the conclusion by \citet{zennarol2019a} that GCs with  multiple populations and $R_{\rm PER}>3.5$ kpc (red starred symbols of Figure\,\ref{fig:2Gmass}) have systematically larger fractions of 1G stars than the remaining Milky Way GCs.
The difference is more extreme in the studied Magellanic Cloud clusters (blue symbols in Figure~\ref{fig:2Gmass}), where the fraction of 1G stars is even larger than that of Galactic GCs with large perigalactic radii.   

We find that the fraction of 1G stars exhibits a weak anti-correlation with the present-day mass of 1G stars as shown lower-left of Figure\,\ref{fig:2Gmass} (R$_{\rm S}$(MW)=$-$0.40$\pm$0.11). In contrast, there is  strong anti-correlation with the present-day mass of 2G stars corresponding to R$_{\rm S}$(MW)=$-$0.77$\pm$0.06 (lower-right panel of Figure\,\ref{fig:2Gmass}). 

Finally, in the upper-right panel of Figure~\ref{fig:2Gmass} we plotted the fraction of 1G stars as a function of the logarithm of the initial cluster masses from \citet{baumgardt2018a} and \citet{goudfrooij2014a}. Initial masses provide stronger correlations with the fraction of 1G stars than those derived from present-day masses, as indicated by the high absolute values of the correlation coefficients, R$_{\rm s}$(MW)$=-0.74 \pm 0.06$,  for Galactic GCs with multiple populations and R$_{\rm s}$(MW)=$-0.82 \pm 0.05$ for all Galactic GCs, respectively.   
 In this case, clusters with large perigalactic radii follow the same relation as GCs with  $R_{\rm PER} \lesssim 3.5$ kpc as demonstrated by the fact that the least-squares best-fit lines obtained from these two groups of clusters overlap each other. 
 
 The fractions of 1G stars in three out of five MC clusters with MPs are comparable with those of Galactic GCs with similar masses at formation, thus supporting the possibility of a universal relation between initial mass and fraction of 1G stars.   
 
 In the uppermost panels of Figure~\ref{fig:2Gmass} we show the histogram distributions of the present-day masses and of the initial masses  (taken at face values from \citet{baumgardt2018a} and \citet{goudfrooij2014a}, but see the caveats discussed in Section 2.2.) for Galactic GCs with multiple populations (gray histograms), Galactic GCs with no evidence of multiple populations (black open histogram), Magellanic Cloud GCs with multiple populations (blue-shaded histogram) and for Magellanic Cloud GCs where we did not detect multiple populations (white-blue histogram). 
 Clearly, both Galactic and MC GCs without multiple populations exhibit, on average, lower present-day masses 
   than MP GCs but the present-day mass distributions of GCs with and without MPs overlap each other. 
The mass difference between simple- and multiple-population Galactic GCs is more pronounced when we consider the initial masses.  Specifically, the fact that all Galactic GCs with initial masses larger than $\sim 1.5 \cdot 10^{5} \mathcal{M}_{\odot}$ host multiple populations, whereas simple-populations Galactic GCs have initial masses, $M_{\rm ini}\lesssim 1.5 \cdot 10^{5} \mathcal{M}_{\odot}$ suggests a possible mass threshold for the formation of MPs \citep[e.g.][]{bragaglia2012a}. 

However, the presence of a universal mass threshold of $\sim1.5 \cdot 10^{5} \mathcal{M}_{\odot}$ for the onset of multiple populations seems  challenged by the evidence that some Magellanic Cloud GCs with initial masses of $\sim 3-4 \cdot 10^{5} \mathcal{M}_{\odot}$ are consistent with simple populations, whereas NGC\,339, with an initial mass of $\sim 1.4 \cdot 10^{5} \mathcal{M}_{\odot}$ hosts MPs.
 We further note that the masses of simple stellar population MC GCs are similar to the masses of 1G stars in multiple-population GCs.

In an attempt to interpret the results of Figure~\ref{fig:2Gmass}, we ran a large number of Monte-Carlo simulations. In each of them, we simulated 10,000 clusters with a  distribution of initial masses that is similar to the one derived by \citet{baumgardt2018a} for the analyzed Galactic GCs, and with an initial fraction of 1G stars, ($N_{\rm 1G}/N_{\rm TOT}$)$_{\rm ini}$, that linearly increases from $f_{\rm 1G, min}$ at $\log{\mathcal{M}_{\rm ini}/\mathcal{M}_{\odot}}=7.0$ to $f_{\rm 1G, max}$ at $\log{\mathcal{M}_{\rm ini}/\mathcal{M}_{\odot}}=5.2$.
 We assumed that each GC retains a fraction of its initial mass of 1G (2G) corresponding to a fixed value $f_{\rm 1G}$ ($f_{\rm 2G}$) plus a Gaussian scatter, $\sigma_{f1G}$ ($\sigma_{f2G}$). Finally, we calculated the resulting fraction of 1G stars, the cluster mass and the total mass of 1G and 2G stars and added to these value the same uncertainties associated with the observations.

\begin{centering} 
\begin{figure*} 
  \includegraphics[width=9.cm,trim={.4cm 5.8cm 0cm 4.0cm},clip]{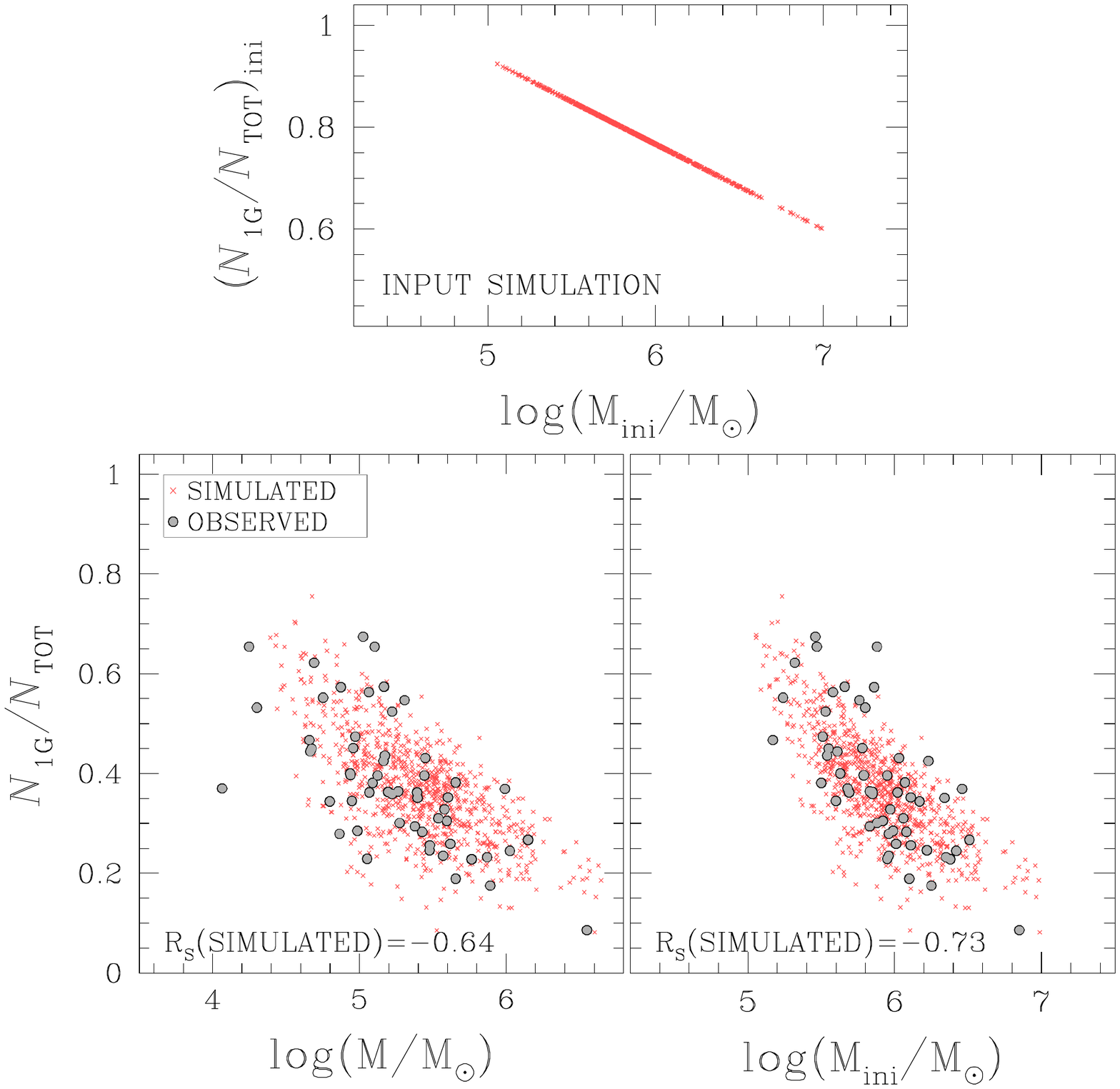}
  \includegraphics[width=9.cm,trim={.4cm 5.8cm 0cm 12.4cm},clip]{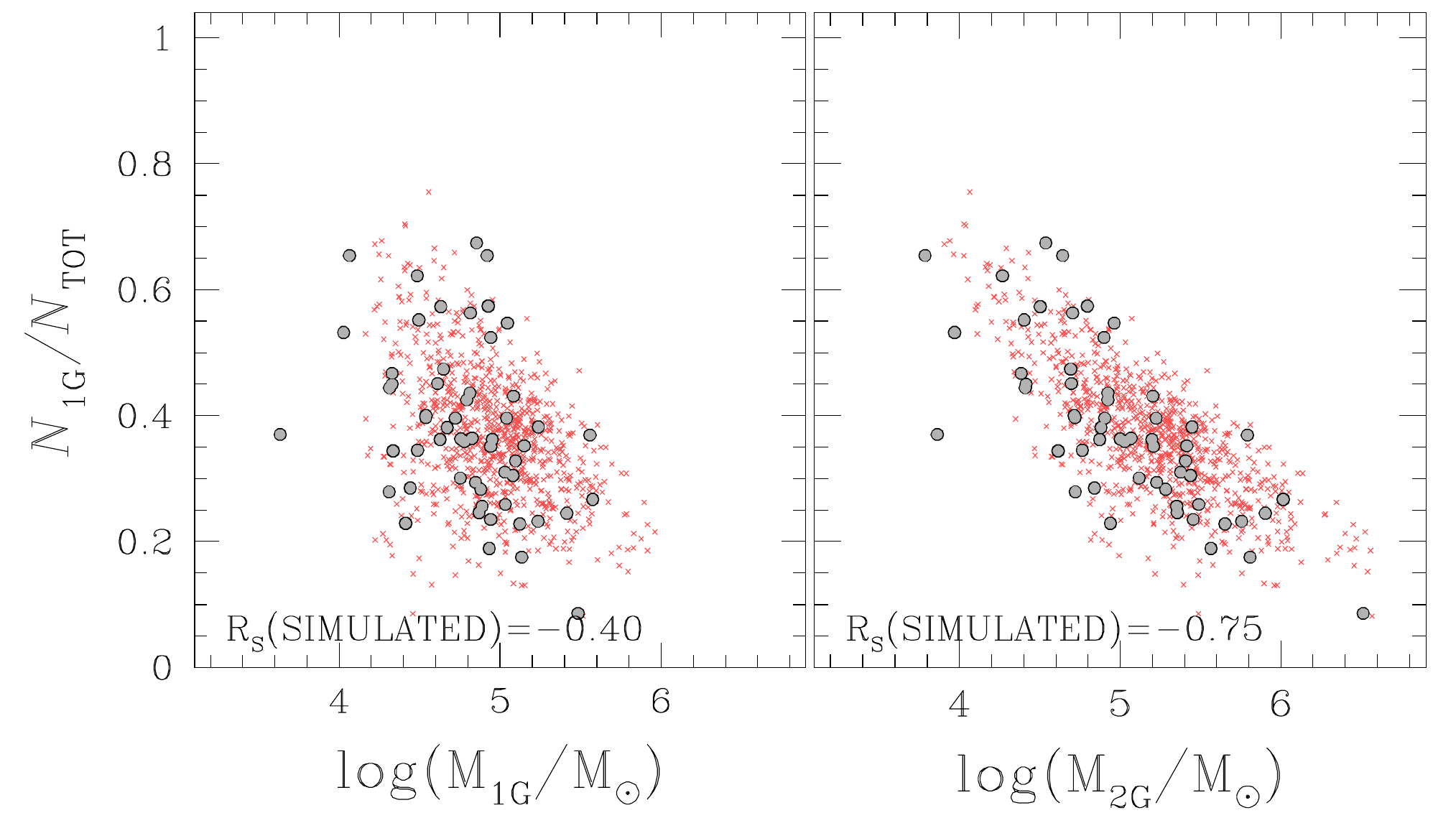}
  \caption{These panels summarize the outcomes from the simulation that best reproduces the results shown in Figure~\ref{fig:2Gmass}. Upper panels show the input fractions of 1G stars against initial cluster masses. 
   We plotted in the other panels the expected present-day fraction of 1G stars against the present-day GC masses (middle-left), the initial GC masses (middle-right) and the present-day total mass of 1G (lower-left) and 2G stars (lower-right). Simulated stars are represented with red crosses, while the gray dots show the observations.
    }
 \label{fig:simu} 
\end{figure*} 
\end{centering}

 The best match with the observations of Figure~\ref{fig:2Gmass} corresponds to $f_{\rm 1G, min}=0.60$ and $f_{\rm 1G, max}=0.88$. In this simulation we assumed that the 1G retains $f_{\rm 1G}=$16\% of its initial mass  in stars with $\sigma_{\rm f1G}=0.04$, while the 2G retains 95\% of its stars, with $\sigma_{\rm f2G}=0.00$. Results from the best-fit simulation are illustrated in Figure~\ref{fig:simu} and indicate that observations are consistent with a scenario
where the stars lost by GCs mostly belong to the 1G.

\subsection{Multiple populations and cluster age}
\begin{centering} 
\begin{figure} 
  \includegraphics[height=8.2cm,trim={.5cm 5cm 0cm 4.5cm},clip]{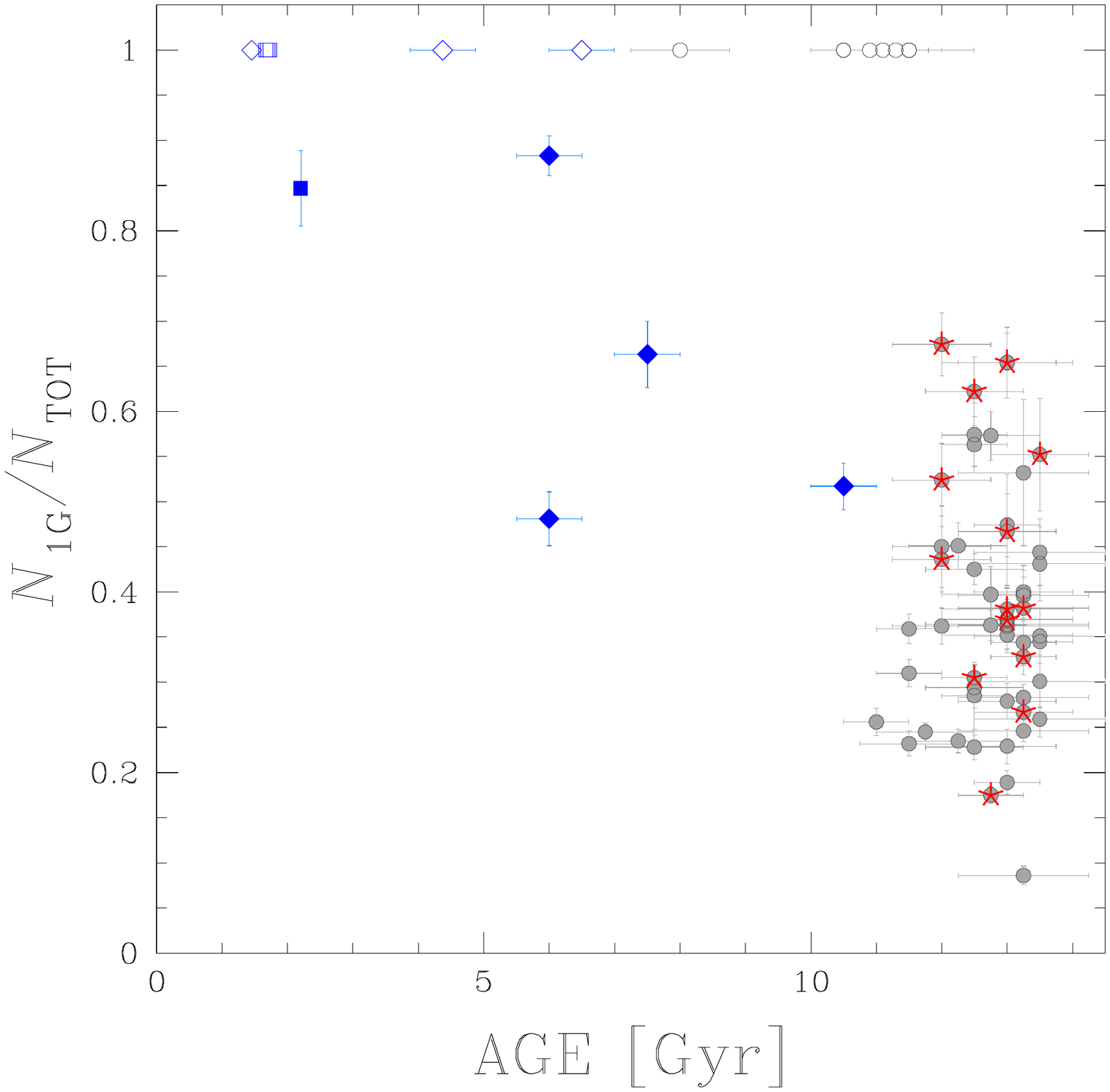}
  \caption{Fraction of 1G stars against cluster ages from \citet{dotter2010a}. 
  Symbols are adopted from Figure\,\ref{fig:2Gmass}.}
 \label{fig:2gVSage} 
\end{figure} 
\end{centering}

Figure\,\ref{fig:2gVSage} shows that there is no evidence for a significant correlation between the fraction of 1G stars in Milky-Way GCs and the cluster ages, although simple-population Galactic GCs have ages between $\sim$8 and 11 Gyr and are younger than the remaining GCs with multiple populations.

 The five LMC and SMC clusters where we detected multiple populations have ages of $\sim$2.0--10.5 Gyr and are younger than the bulk of Milky Way GCs. Their fractions of 1G stars are, on average, higher than those of Galactic GCs, although NGC\,121, NGC\,416 and NGC\,339 host similar fractions of 1G stars as some $\sim$13 Gyr-old Milky Way clusters. 
 
 As discussed in the previous subsection, the fraction of 1G stars mostly depends on the mass of the host GC, and that it is challenging to compare masses of clusters with different ages of different galaxies. As a consequence, we believe that it is not possible to draw any strong conclusion on a possible relation between the fraction of 1G stars and cluster age, without properly removing the effect of cluster mass \citep[see also discussion by][]{lagioia2019b}.     

\subsection{Multiple populations and helium abundance}
\begin{centering} 
\begin{figure} 
  \includegraphics[height=8.2cm,trim={.8cm 5.cm 0.15cm 4.5cm},clip]{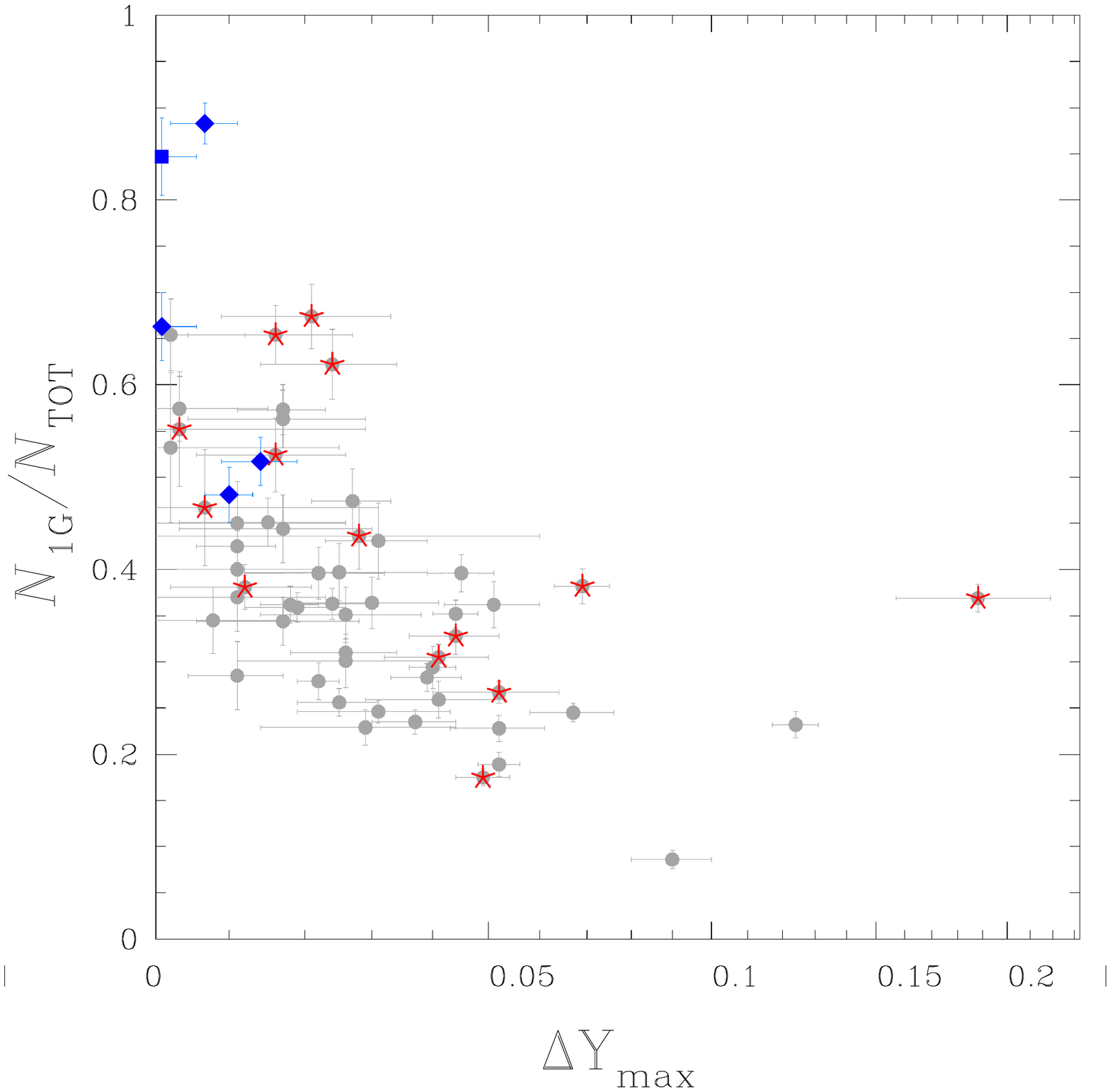}
  \caption{Fraction of 1G stars as a function of the maximum internal helium variations \citep[][]{milone2018a, zennarol2019a, lagioia2019a}. See Figure~\ref{fig:2Gmass} for symbol definition.}
 \label{fig:2gVSdy} 
\end{figure} 
\end{centering}

The maximum internal helium variation in GCs ranges from less than $0.01$ to more than $0.15$ in mass fraction and correlates with the total luminosity and the present-day mass of the host GC \citep[e.g.][]{milone2015b, milone2018a, lagioia2018a, zennarol2019a}.
Figure~\ref{fig:2gVSdy} shows that clusters dominated by 1G stars exhibit small helium spreads and the fraction of 1G stars anti-correlates with the maximum helium variation.  The fraction of 1G stars rapidly increases in GCs with $\Delta$ Y$_{\rm max} \lesssim 0.03$, while for the clusters with large helium variations the 1G fraction generally lies between $\sim$0.1 and 0.4.

\begin{centering} 
\begin{figure*} 
  \includegraphics[width=12.cm,trim={.4cm 5.8cm 0cm 11.0cm},clip]{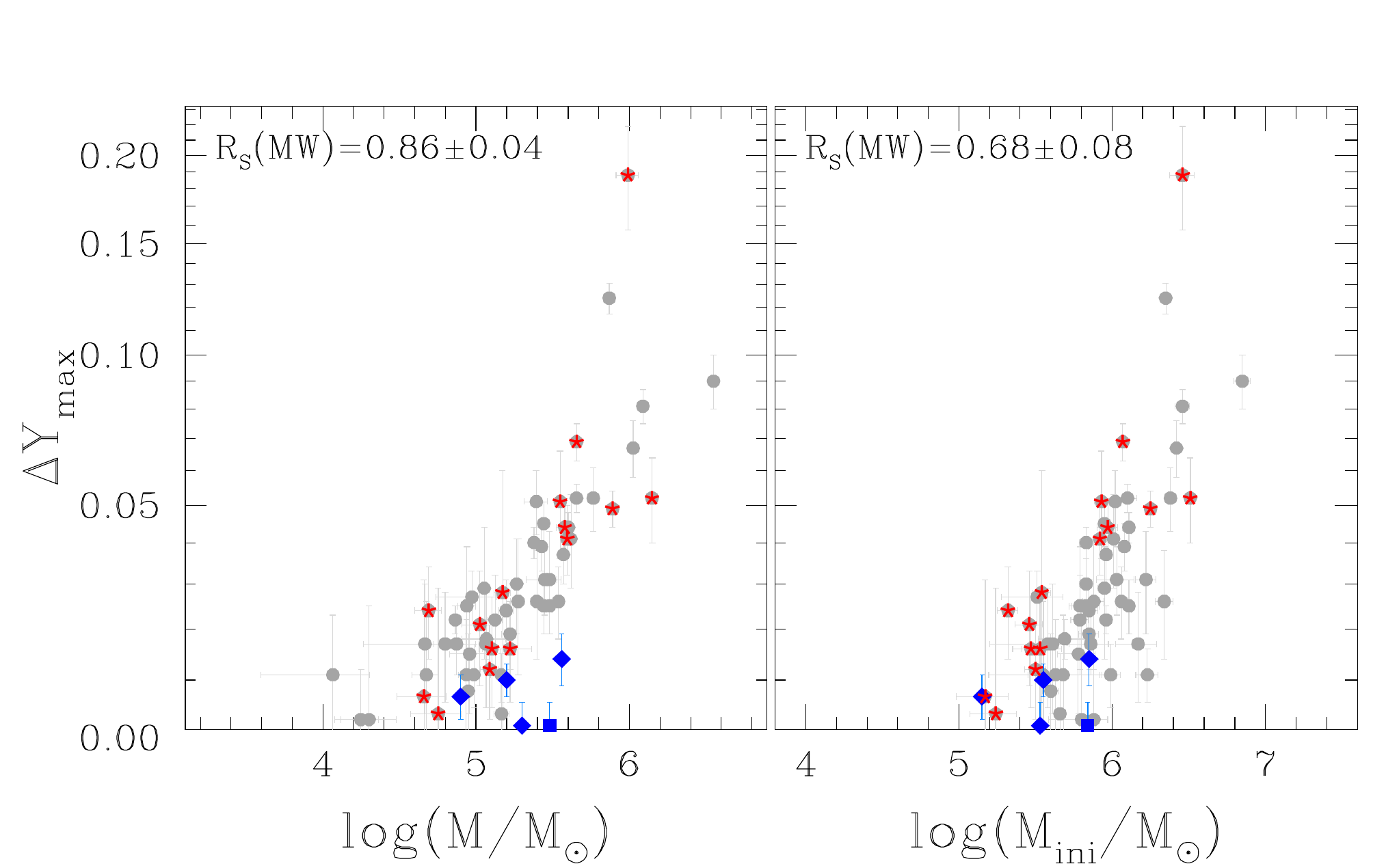}
  \includegraphics[width=12.cm,trim={.4cm 5.8cm 0cm 12.4cm},clip]{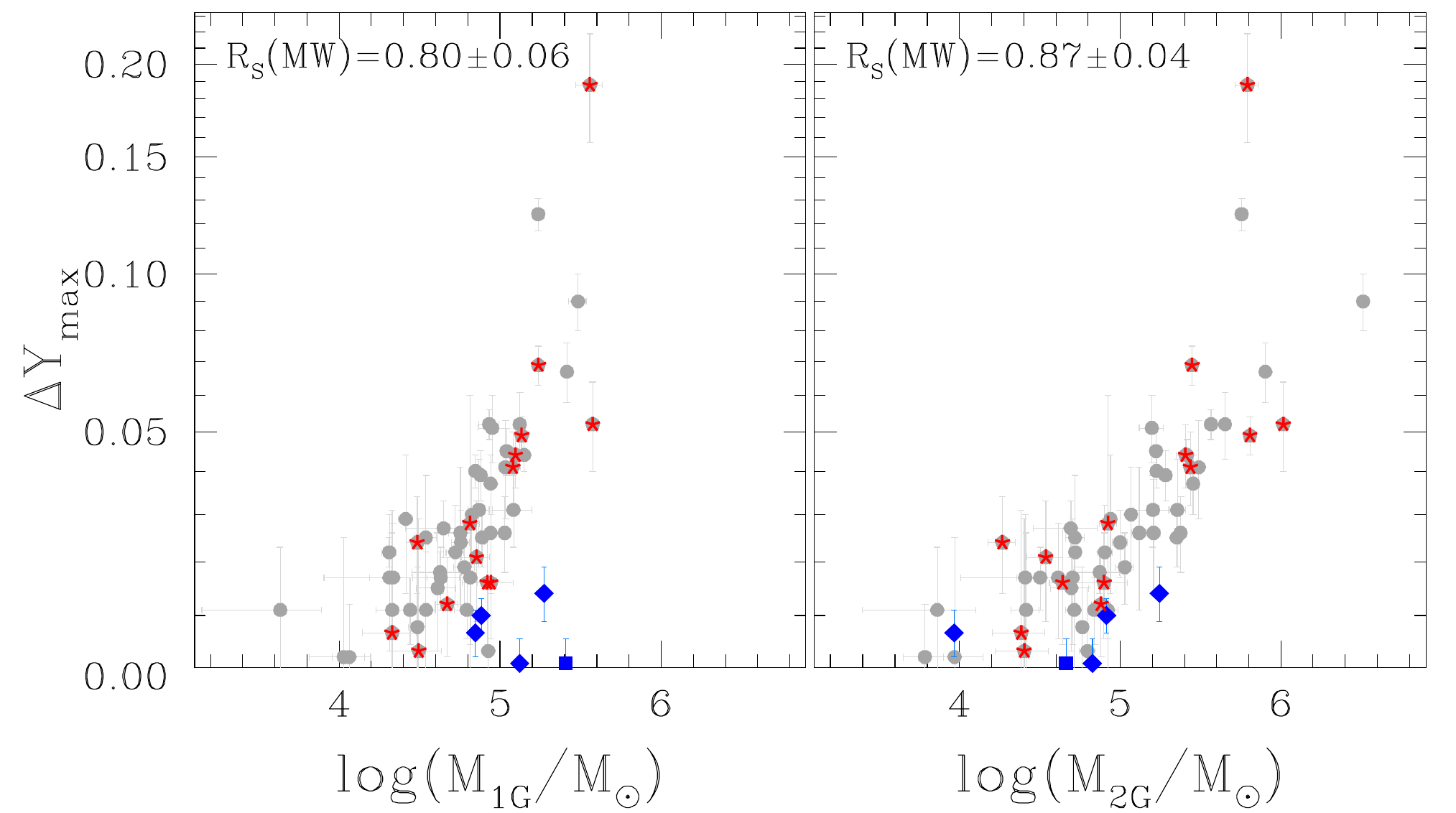}
  \caption{Maximum internal helium variation against the present-day cluster mass (upper-left) the initial cluster mass (upper-right) for Galactic GCs, the present-day mass of the 1G (lower-left) and the present-day mass of the 2G (lower-right). Symbols are like in Figure~\ref{fig:2Gmass} and only GCs with MPs are plotted.}
 \label{fig:dymax} 
\end{figure*} 
\end{centering}

Figure~\ref{fig:dymax} shows that the maximum internal helium variation of Galactic GCs correlates with the present day cluster mass and that Galactic GCs with $R_{\rm PER} >3.5$ kpc exhibit the same trend as the remaining Galactic clusters.  
As regards the five Magellanic Cloud clusters with multiple populations, the range in $\Delta$Y$_{\rm max}$ is small: a correlation with present-day mass may exist, but relative to the Galactic GCs, the $\Delta$Y$_{\rm max}$ values at constant present-day mass are smaller.
The correlation between the maximum helium variation and the logarithm of cluster mass is still present when we use the initial masses, but the significance of the correlation for Milky Way GCs is $R_{\rm S}$(MW)=0.68$\pm$0.08 and thus is lower than that obtained from present-day masses ($R_{\rm S}$(MW)=0.86$\pm$0.04). For a fixed initial mass, Galactic GCs with large perigalactic radii exhibit, on average, larger internal helium spreads than the remaining clusters. MC clusters follow the same trend as Galactic GCs in the  $\Delta$Y$_{\rm max}$ vs.\,$\log{\mathcal{M}_{\rm ini}/\mathcal{M}_{\sun}}$ plane.

$\Delta$Y$_{\rm max}$ correlates with the present-day masses of both 1G and 2G stars, with the latter showing the highest values of the Spearman's correlation rank coefficient ($R_{\rm S}$(MW)=0.87$\pm$0.04).
Magellanic Cloud clusters have lower values of $\Delta$Y$_{\rm max}$ than Galactic GCs with similar 1G masses and the difference diminishes when we use the masses of 2G stars.
 \begin{table*}
  \caption{Properties of the LMC and SMC clusters studied in this paper.
   }
\begin{tabular}{l r r c c c c c}
\hline \hline
ID & age [Gyr] & $R_{\rm eff}$ [pc]  & $\log{\mathcal{M}/\mathcal{M_{\odot}}}$  & $\log{\mathcal{M_{\rm ini}}/\mathcal{M_{\odot}}}$ & $\log{\mathcal{M_{\rm ini}^{\rm seg}}/\mathcal{M_{\odot}}}$ & $N_{\rm 1G}/N_{\rm TOT}$ & $\Delta$Y$_{\rm max}$ \\
\hline
Lindsay\,1    & 8.0$\pm$0.5$^{b}$ & 17.23$\pm$1.61$^{e}$ & 5.30$^{g}$          & 5.53$^{a}$ & 6.23$^{a}$ & 0.663$\pm$0.037$^{a}$ & 0.000$\pm$0.004$^{b}$\\
Lindsay\,38   & 6.0$\pm$0.5$^{c}$ & 12.74$\pm$0.53$^{e}$ & 4.70$^{g}$          & 5.13$^{a}$ & 5.79$^{a}$ & 1.00$^{a}$            & 0.000$^{a}$ \\
Lindsay\,113  & 4.5$\pm$0.5$^{c}$ & 15.25$\pm$1.38$^{a}$ & 4.36$^{h}$          & 4.61$^{a}$ & 5.10$^{a}$ & 1.00$^{a}$            & 0.000$^{a}$ \\
NGC\,121      &10.5$\pm$0.5$^{b}$ & 8.50$\pm$0.70$^{e}$  & 5.56$^{g}$          & 5.85$^{a}$ & 6.31$^{a}$ & 0.517$\pm$0.026$^{a}$ & 0.014$\pm$0.004$^{a}$ \\
NGC\,339      & 6.5$\pm$0.5$^{b}$ & 11.69$\pm$0.66$^{e}$ & 4.90$^{g}$          & 5.15$^{a}$ & 5.81$^{a}$ & 0.883$\pm$0.022$^{a}$ & 0.007$\pm$0.004$^{b}$ \\
NGC\,416      & 6.0$\pm$0.5$^{b}$ & 4.97$\pm$0.77$^{e}$  & 5.20$^{g}$          & 5.55$^{a}$ & 6.05$^{a}$ & 0.481$\pm$0.030$^{a}$ & 0.010$\pm$0.003$^{b}$ \\
NGC\,419      & 1.6$\pm$0.1$^{c}$ & 5.38$\pm$0.08$^{f}$  & 5.38$^{f}$          & 5.51$^{f}$ & 5.94$^{f}$ & 1.00$^{a}$            & 0.000$^{a}$ \\
NGC\,1783     & 1.6$\pm$0.1$^{d}$ & 5.42$\pm$0.11$^{f}$  & 5.42$^{f}$          & 5.54$^{f}$ & 5.98$^{f}$ & 1.00$^{a}$            & 0.000$^{a}$ \\
NGC\,1806     & 1.6$\pm$0.1$^{d}$ & 5.10$\pm$0.06$^{f}$  & 5.10$^{f}$          & 5.23$^{f}$ & 5.66$^{f}$ & 1.00$^{a}$            & 0.000$^{a}$ \\
NGC\,1846     & 1.6$\pm$0.1$^{d}$ & 5.24$\pm$0.09$^{f}$  & 5.24$^{f}$          & 5.37$^{f}$ & 5.80$^{f}$ & 1.00$^{a}$            & 0.000$^{a}$ \\
NGC\,1978     & 2.0$\pm$0.1$^{d}$ & 6.81$\pm$0.19$^{a}$  & 5.48$^{a}$          & 5.84$^{a}$ & 6.28$^{a}$ & 0.847$\pm$0.042$^{a}$ & 0.001$\pm$0.003$^{a}$ \\
    \hline\hline
\end{tabular}
\\
  References: $^{a}$ This paper; $^{b}$ \citet{lagioia2019a}; $^{c}$ \citet{glatt2008a};
  $^{d}$\citet{milone2009a};
  $^{e}$ \citet{glatt2009a}; $^{f}$\citet{goudfrooij2014a};
  $^{g}$ \citet{glatt2011a};  $^{h}$ \citet{chantereau2019a}.
  \label{tab:par}
 \end{table*}

\subsection{Multiple populations in Galactic Globular Clusters and the progenitor galaxy}
Recent results, mostly based on data provided by the Gaia mission, revealed that the Milky Way and its GCs have experienced a complex assembly history. 
\citet{massari2019a} analyzed the dynamics of Galactic GCs to identify clusters with common origins. They argue that $\sim$40\% of clusters would have formed in situ, in the Galaxy which they designate as the `Main Progenitor'.  This group includes 36 Bulge clusters (MB) and 26 disk clusters (MD).  On the other hand, about 35\% of Milky Way GCs  are the results of merger events with from the merger events of the Gaia-Enceladus dwarf galaxy  \citep[GE,][]{helmi2018a}, of the progenitor of the Helmi streams \citep[H99,][]{helmi1999a, koppelman2019a}, of the Sagittarius dwarf galaxy \citep[Sag,][]{ibata1994a} and of the Sequoia galaxy \citep[Seq,][]{myeong2019a}.  As for the other clusters, about 16\% appear to be associated with a group of high-binding energy clusters (LE) with the remainder on loosely bound orbits (HE) with heterogeneous origins.

\begin{centering} 
\begin{figure*} 
  \includegraphics[height=6.4cm,trim={0cm 5cm .5cm 4.5cm},clip]{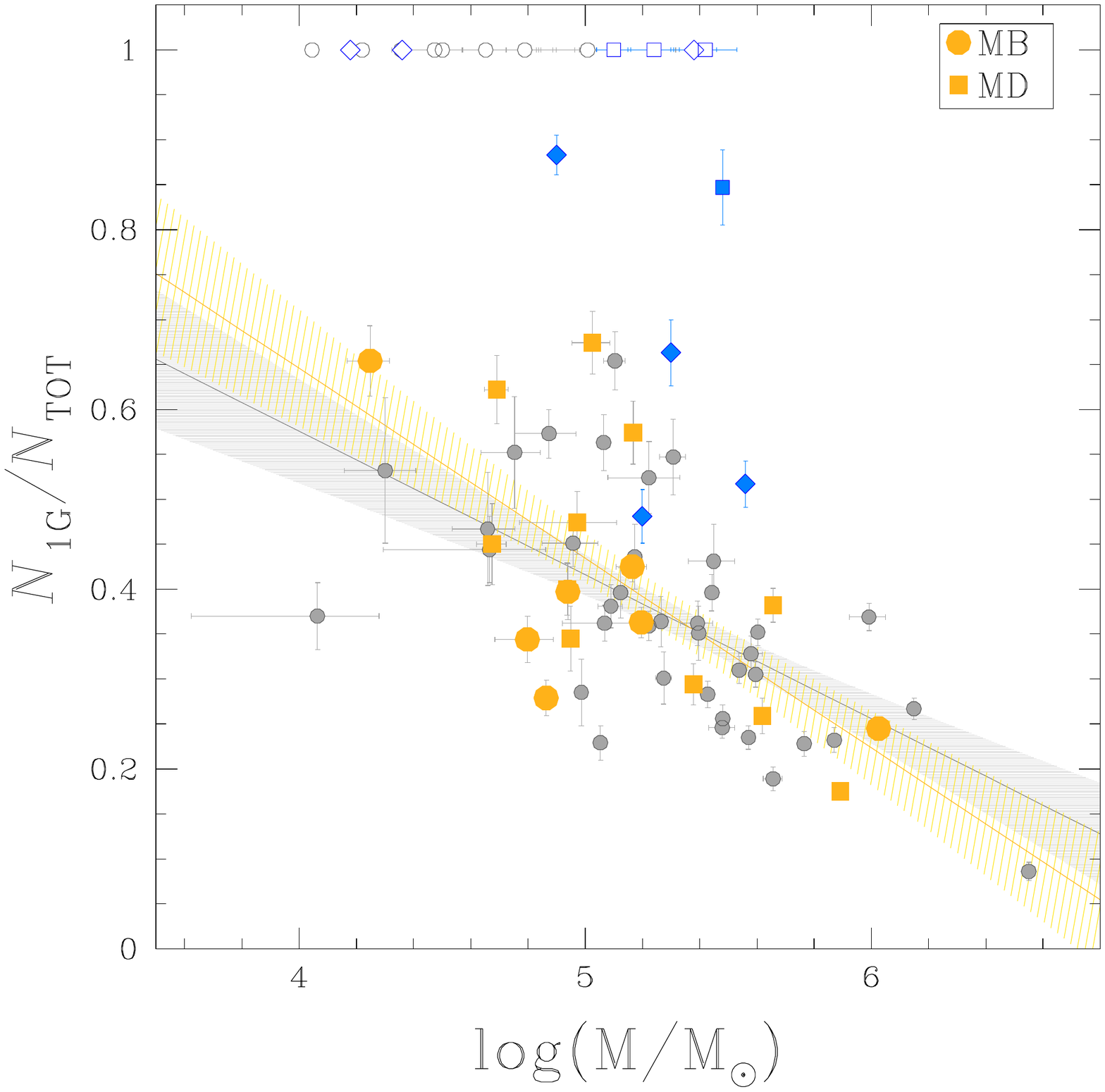}
  \includegraphics[height=6.4cm,trim={3.4cm 5cm .5cm 4.5cm},clip]{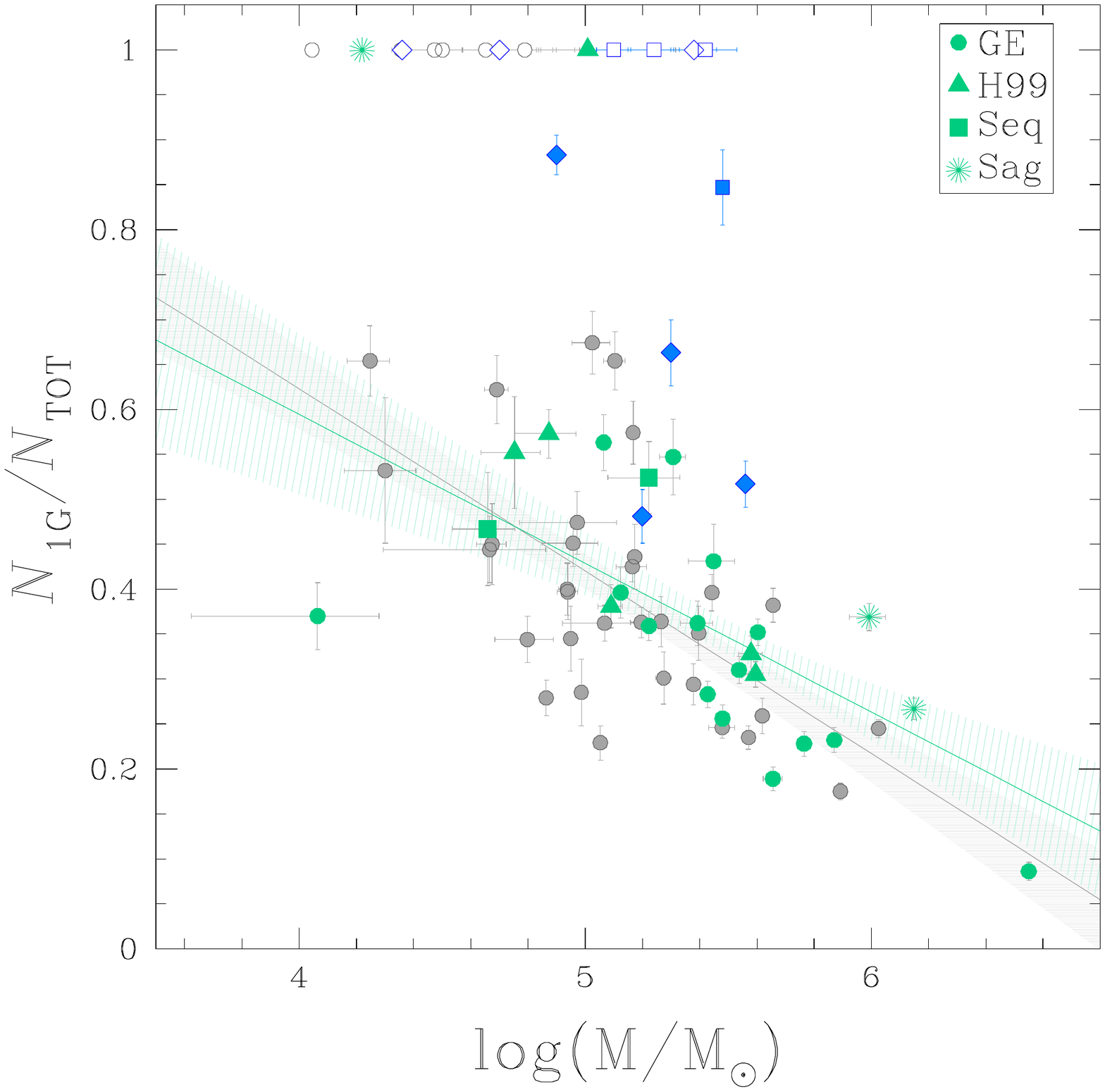}
  \includegraphics[height=6.4cm,trim={3.4cm 5cm .5cm 4.5cm},clip]{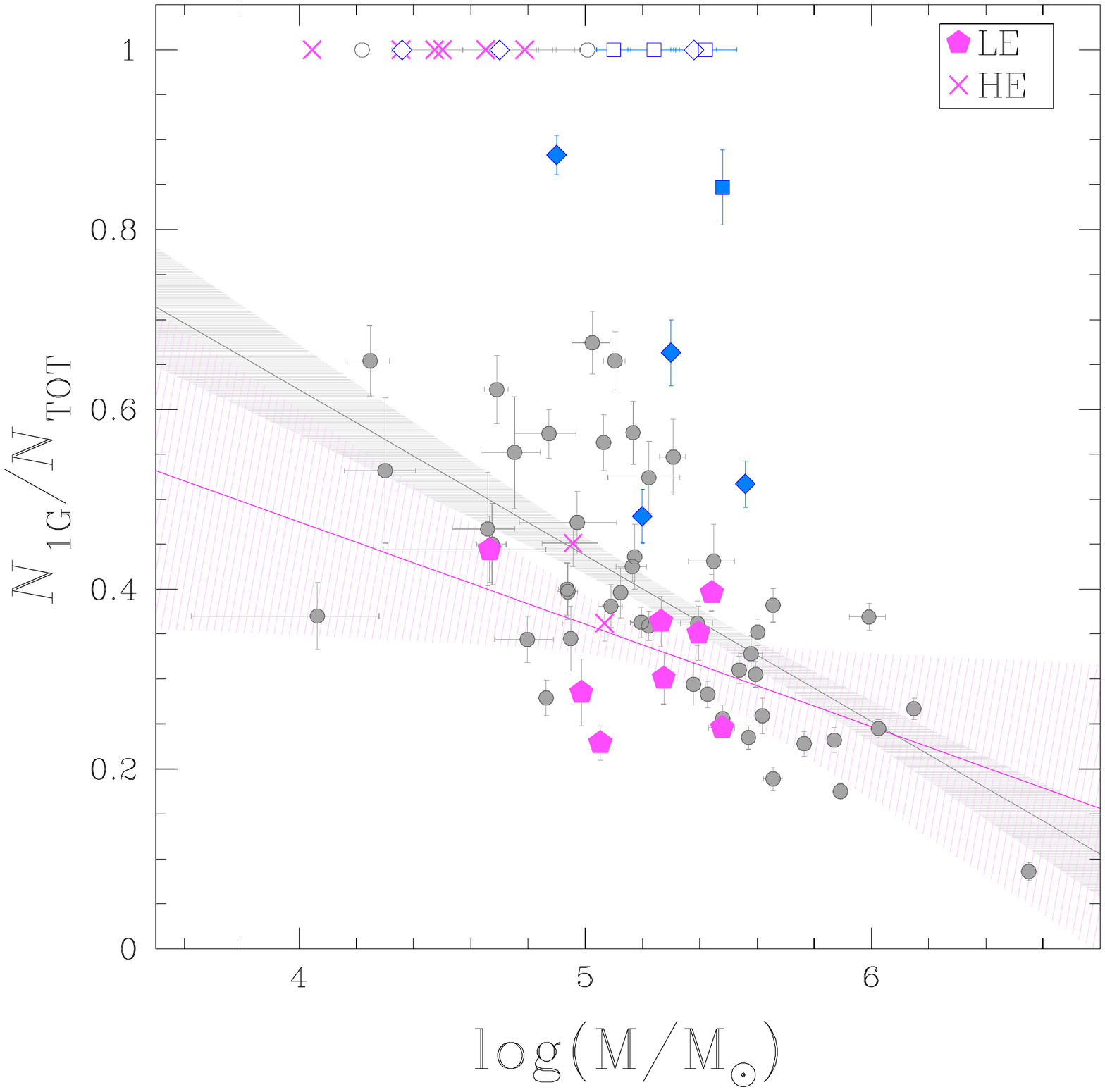}
  \caption{Each panel is a reproduction of the upper-left panel of Figure\,\ref{fig:2Gmass} where we plot the fraction of 1G stars against the present-day mass of the host GC. 
  Blue colors indicate LMC (squares) and SMC clusters (diamonds). Simple- and multiple-population GCs are indicated with open and filled symbols, respectively.
   In the left, middle, and right panels we used yellow, aqua and magenta colors to plot Milky-Way GCs formed in-situ, GCs that are the results of mergers, and other GCs with either high or low energy, respectively. 
      The different colored symbols indicate the progenitor structure according to \citet{massari2019a}. GCs not in the specific categories are plotted as gray symbols. 
      The corresponding least-square best fit lines are represented with the same colors and the shaded areas include the 68.27\% confidence interval of the true regression.
      }
 \label{fig:2Gmass2} 
\end{figure*} 
\end{centering}

We now use the allocations by \citet{massari2019a} to investigate the properties of GCs with different progenitors. In the left and middle panels of Figure\,\ref{fig:2Gmass2} we show that clusters formed in situ (yellow symbols), clusters that result from mergers (aqua symbols), and LE clusters follow very-similar behaviours in the $N_{\rm 1G}/N_{\rm TOT}$ vs.\,$log{\mathcal{M}/\mathcal{M}_{\odot}}$ plane. 
This result suggests that there is no evidence for a significant difference between the analyzed behaviour of multiple populations and the tentative progenitor galaxy.  
Possible exceptions include the facts that some LE clusters have low 1G fractions compared to other GCs of similar mass, and that all unassociated HE GCs but NGC\,6584 and NGC\,6934 are consistent with simple populations.

Historically, GCs have been tentatively grouped in different subsystems on the basis of their metallicities and horizontal-branch morphologies alone, \citep[e.g.][]{zinn1993a, vandenberg1993a}.
\citet{mackey2004a} defined three subsystems of `young' halo GCs, possibly formed in external satellite galaxies, `old' halo GCs and Bulge-Disk GCs, which are believed to be born in the Milky Way. For completeness, we applied the same analysis discussed above to those three groups of GCs defined by Mackey and Gilmore and find that they share similar behaviours, 
 thus corroborating the conclusion that the fraction of 1G stars does not significantly depend on the candidate host galaxy. 

\subsection{Type II GCs in the integral of motion space}
In recent papers, \citet{milone2017a} and \citet{marino2019a} defined 
defined two classes of GCs, designated Type I and Type II, based on the ChMs of 58 clusters.  In this sample, $\sim$83\% exhibit a single sequence of 1G and 2G stars in the ChM; these are the Type I clusters.  The Type II clusters show multiple sequences of 1G and 2G stars in the ChMs and optical band photometry of these clusters reveals split SGBs in the CMD \citep[e.g.][]{milone2008a, marino2009a, piotto2012a}.
Studies based on the synergy of photometry and spectroscopy revealed that Type II GCs correspond to the class of `anomalous' GCs with star-to-star variations in some heavy elements, like Fe and s-process elements. The sample of all Type II GCs  comprises NGC\,362, NGC\,1261, NGC\,1851, $\omega$\,Centauri, NGC\,5286, NGC\,6273, NGC\,6388, NGC\,6656 (M\,22), NGC\,6715 (M\,54), NGC\,6934, NGC\,7078 (M\,15), NGC\,7089 (M\,2) and Terzan\,5
\citep[e.g.][]{marino2009a, marino2015a, marino2018c, dacosta2009a, yong2008a, yong2014a, yong2016a, carretta2010a, johnson2015a, johnson2017a, nardiello2018a, ferraro2009a, massari2014a}. 

Due to their complex chemical composition, Type II GCs have been often associated with remnants of dwarf galaxies that have been cannibalized by the Milky Way \citep[e.g.][]{marino2015a, marino2017a}. This possibility is supported by the fact that this class of clusters includes M\,54, in the nucleus of the Sagittarius dwarf spheroidal galaxy, and $\omega$\,Centauri, which is the most-massive Milky Way GC and, due to the extreme metallicity variation, has been considered as the surviving remnant of a tidally disrupted dwarf galaxy.

\begin{centering} 
\begin{figure*} 
  \includegraphics[height=7.5cm,trim={0.5cm 5cm 6.2cm 10cm},clip]{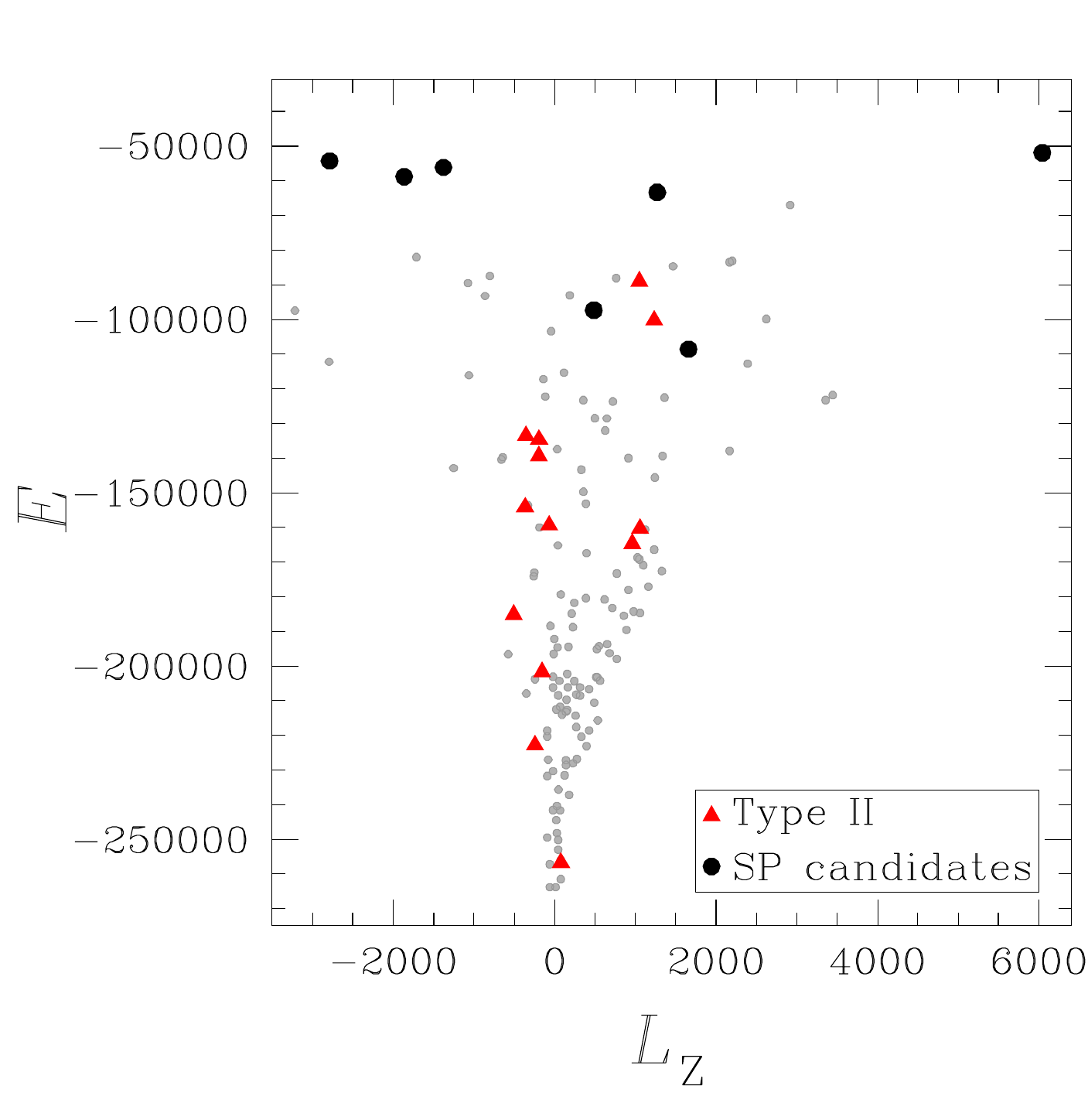}
  \includegraphics[height=7.5cm,trim={0.5cm 5cm 0.2cm 10cm},clip]{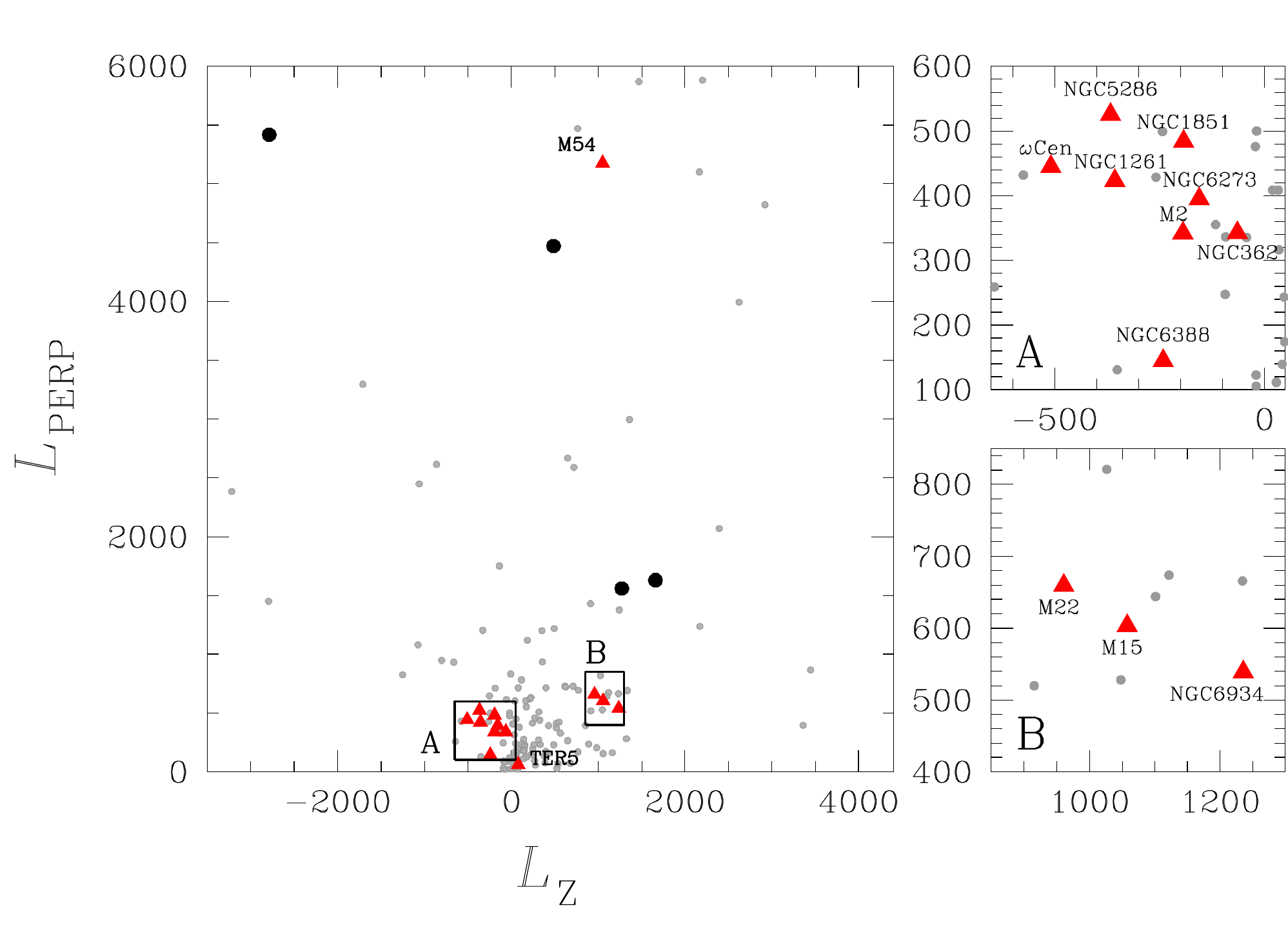}
  \caption{$E$ vs.\,$L_{\rm Z}$ (left panel) and $L_{\rm PERP}$ vs.\,$L_{\rm Z}$ (right panel) projections of integral of motion space for Galactic GCs \citep[from][]{massari2019a}. Type II GCs are marked with red triangles, while candidate simple-population (SP) GCs are indicated with black dots. In the left panel we only show  GCs with $-3,500 <L_{\rm Z}<6,400$ m/s kpc, thus excluding the simple population cluster AM\,1, which has $L_{\rm Z} \sim -12,400$ km/s kpc and E$\sim -42,000$ km$^{2}$/s$^{2}$ \citep{massari2019a}.
   In the right panel we show only GCs with $L_{\rm PERP}<6,000$ m/s kpc and with  $-3,500 <L_{\rm Z}<4,400$ m/s kpc. The small panels on the right are zoom in of the two regions A and B of the $L_{\rm PERP}$ vs.\,$L_{\rm Z}$ diagram that include most Type II GCs. }
 \label{fig:ene} 
\end{figure*} 
\end{centering}

Figure\,\ref{fig:ene} reveals that most Type II GCs are clustered in two distinct regions of the integral of motion space \citep[from][]{massari2019a}. 
We distinguish a main group of seven Type II clusters, including $\omega$~Cen, NGC\,362, NGC\,1261, NGC\,1851, NGC\,5286, NGC\,6273 and NGC\,7089 (M\,2), with $-500 \lesssim L_{\rm Z} \lesssim 0$ km/s kpc and $300 \lesssim L_{\rm PERP} \lesssim 550$ km/s kpc, shown as region A in Figure~\ref{fig:ene}. NGC\,6388 shares similar values of $L_{\rm PERP}$ as these seven clusters but smaller $L_{\rm PERP} \sim 150$ km/s kpc.

Based on 100,000 Monte-Carlo simulations, where we assumed that the simulated GCs have the same distribution in the IOM as the observed clusters, we find that the probability that seven (or more) out of thirteen randomly-extracted GCs populate a 500$\times$250 (km/s kpc)$^{2}$ square region of the $L_{\rm PERP}$ vs.\,$L_{\rm Z}$ plane is 0.021. 
In all these cases, the area that includes the seven GCs is centered in the region with $L_{\rm PERP} <300$ km/s kpc, where we observe the highest density of clusters. 
Hence, we conclude that 
 the fact that seven Type II GCs are clustered in such a small region of the $L_{\rm PERP}$ vs.\,$L_{\rm Z}$ plane 
 around  $L_{\rm Z}\sim -250$ km/s kpc and $L_{\rm PERP} \sim330$ km/s kpc  is unlikely a coincidence due to random event. 
 As a consequence, at least these seven clusters are possibly associated with the same parent galaxy. This is consistent with the associations given in \citet{massari2019a}, who found six out of these seven GCs to be likely linked to Gaia-Enceladus. The only remaining cluster, namely NGC\,6723, was associated to the LE group, that is located next to Gaia Enceladus in the IOM space. Given the uncertainty in the associations of clusters to progenitors with similar dynamical properties, it is not unreasonable to consider NGC\,6723 as well as a possible member of Gaia Enceladus.
 
We also note that four Type II GCs have $-900 \lesssim L_{\rm Z} \lesssim 1250$ km/s kpc and  $-170,000 \lesssim E \lesssim -80,000$ km$^{2}$/s$^{2}$. Three of them, NGC\,6656 (M\,22), NGC\,6934 and NGC\,7078 (M\,15), populate the region B of the $L_{\rm PERP}$ vs.\,$L_{\rm Z}$ plane with similar values of $L_{\rm Z}$ between $\sim$500 and 700 km/s kpc. The fourth cluster is M\,54, whixh has a much higher value of $L_{\rm perp}$.
However, the small number of three GCs, prevents us from any conclusion on their origin. 
For completeness, we show that the sample of high-energy GCs  is  mostly composed of simple-population GCs (large black dots in Figure~\ref{fig:ene}).

\section{Summary and final remarks}\label{sec:summary}

In this work we compare the properties of MPs in Magellanic Cloud GCs and in Galactic GCs either formed in situ or associated with various progenitor galaxies. Our goal is to investigate whether the presence  of MPs is a universal phenomenon or depends on the host galaxy. The main results of the paper can be summarized as follows:
\begin{itemize}
\item
We derived the $\Delta_{{\it C} \rm F336W,F343N,F438W}$ vs.\,$\Delta_{\rm F438W,F814W}$ ChM of eleven GCs in the LMC and SMC with ages between $\sim$1.5 and $\sim$10.5 Gyr to search for evidence of MPs with different light-element abundances.
We find that the ChMs of Lindsay\,1, NGC\,121, NGC\,339, NGC\,416 and NGC\,1978 are not consistent with a simple population thus confirming previous results based on CMDs made with photometry in appropriate filters \citep[e.g.][]{dalessandro2016a, niederhofer2017a, lagioia2019b, martocchia2018a, martocchia2018b, martocchia2019a, chantereau2019a} and on stellar spectroscopy \citep[][]{hollyhead2017a, hollyhead2018a}.  

\item
From the ChMs we estimated the fractions of 1G and 2G stars in these MC clusters, in the same manner as previously done for Milky Way GCs.
We find that the fraction of 2G stars ranges from $\sim$15\% in NGC\,339 and NGC\,1978 to about 50\% in NGC\,121. The remaining GCs, Lindsay 38, NGC\,419, Lindsay 113, NGC\,1783, NGC\,1806 and NGC\,1846, show no evidence of MPs in agreement with the conclusion drawn by \citet{mucciarelli2014a, milone2017b, martocchia2017a, li2019a} for these clusters. 

\item
We compared the population ratios derived for these MC clusters with the fractions of 1G stars measured from the ChMs of 56 Galactic GCs by \citet{milone2017a} and \citet{zennarol2019a}. Moreover, we included in the analysis eight Galactic GCs that are likely composed of a simple population \citep[ but are still lacking a ChM analysis,][]{villanova2013a, milone2014a, dotter2018a, lagioia2019b}. 
\citet{milone2017a} show that the fraction of 1G stars in Galactic GCs ranges from $\sim$10\% to more than 60\%, a quantity that anti-correlates with the total luminosity \citep[from the 2010 version of the][catalog]{harris1996a} and the present-day mass of the host cluster \citep[from][]{mclaughlin2005a}. The same correlation is confirmed when the present-day and initial cluster masses derived by \citet{baumgardt2018a} are used. 

\item
The fraction of 1G stars in the five MC clusters in which MPs are detected also seems to anti-correlate with mass, although the small number prevents firm conclusions.
 Magellanic Cloud clusters with MPs host typically larger fractions of 1G stars than Galactic GCs with similar present-day masses. 

\item 
Simple-population Galactic GCs have initial masses smaller than $\sim 1.5 \cdot 10^{5}$ M$_{\odot}$ thus suggesting that a mass threshold governs the occurrence of MPs. 
This conclusion is challenged by four simple-population MC GCs, namely  NGC\,419, NGC\,1783, NGC\,1806 and NGC\,1846, with initial masses of $\sim 1.5 - 3.5 \cdot 10^{5}$ M$_{\odot}$. 
The fact that these four clusters have ages of $\sim 1.6$\,Gyr is consistent with the conclusion by \citet{bastian2018a} and \citet{martocchia2019a} that MPs could appear only in GCs older than $\sim$2 Gyr.   However, given the uncertainties in the initial mass determinations that we discussed in Section~\ref{sec:data}, we can not exclude that the difference between the initial masses of these young MC and the most massive simple population Galactic GCs, is due to systematic errors in the initial mass estimates of either Galactic GCs or MC clusters,  or both. 

\item
Our analysis reveals that the fraction of 1G stars in all GCs with MPs exhibits a strong anti-correlation with the present-day mass of the 2G but only a mild correlation with the present-day mass of 1G stars. 
When we compare the fraction of 1G stars with the initial cluster masses we also obtain a strong anti-correlation, with a value of the Spearman's rank correlation coefficient that is higher than that obtained from the present-day mass.  Clusters with large and small perigalactic radii share a similar behaviour when the initial masses are used and MC clusters with MPs follow the same trend defined by Galactic GCs. 

These results are consistent with a scenario where the fraction of 1G stars decreases with cluster mass at formation and the GCs have lost a large fraction of 1G stars but a smaller amount of 2G stars. This scenario would result in strong anti-correlations between the fraction of 1G stars and both the initial mass of the host cluster and the present-day mass of 2G stars. We would also expect a less-significant correlation with the present-day mass of 1G stars. As a consequence, the correlation between present-day cluster mass and the fraction of 1G stars would exhibit a lower significance than the corresponding correlation with the initial mass.   
This result is consistent with the predictions by several authors that GCs preferentially lost their 1G stars \citep[e.g.][]{dercole2008a, dercole2010a, dantona2016a}.

\item
The maximum helium abundance variation in the five MC GCs with MPs ranges from $\Delta$Y$_{\rm max} \sim 0.00$ to less than $\Delta$Y$_{\rm max} = 0.02$ and may correlate with the present-day cluster mass, in a similar fashion what is observed in Galactic GCs \citep[][]{milone2018a, zennarol2019a}.  Galactic GCs with different values of the perigalactic radii follow the same behaviour in the $\Delta$Y$_{\rm max}$ vs.\,$\log{\mathcal{M}/\mathcal{M}_{\odot}}$ plane, in contrast with the Magellanic-Cloud GCs, which host smaller helium variations than Galactic GCs with similar masses.

The maximum helium variation in Galactic GCs correlates with the initial cluster mass, but this correlation is less significant (R$_{S}$(MW)=0.68$\pm$0.08) than that observed with the present-day mass (R$_{S}$(MW)=0.86$\pm$0.04). Moreover, Galactic GCs with large perigalactic  radii exhibit larger values of $\Delta$Y$_{\rm max}$ than Milky Way GCs with similar initial masses small perigalactic radii. MC clusters follow the same relation between maximum helium variation and initial masses.
This observational evidence is consistent with a scenario where the helium variation depends on the total mass of the 2G.

\item
Based on the work by \citet{massari2019a}, 
who linked most Galactic GCs to a variety of progenitor systems, we analyzed multiple populations in 17 GCs formed in situ, 25 clusters that are considered the products of merging processes, and 17 other GCs that are not associated with any parent  stellar stream and and which are characterized by either high or low energy. When we plot the fraction of 1G stars against the cluster mass, these three groups of GCs follow nearly the same pattern. As a consequence, there is no evidence for any dependence of the present day population ratio in GCs on the progenitor system. 

\item
Six out of eight candidate simple-population GCs are unassociated high-energy clusters. The remaining two simple-population clusters,  Rup\,106 and Terzan\,7 are associated with the progenitor of the Helmi stream and to the Sagittarius dwarf spheroidal, respectively.
Based on these results, we speculate that simple-population GCs are low-mass clusters that formed in the environment of dwarf galaxies.

\item
The recently identified class of Type~II, or `anomalous' GCs is composed of thirteen known GCs with internal variations of heavy elements.
Our investigation, together with other work, based on high-precision {\it HST} photometry of six additional LMC clusters \citep{Wagner-Kaiser2017a}, reveal that there is no evidence of Type~II GCs in LMC and SMC GC populations. 
To understand whether Type~II GCs have an extragalactic origin, as suggested by the fact that the nuclear star cluster M\,54 of the Sagittarius Dwarf Spheroidal belongs to this class of objects, we investigate their position in the integrals of motions space.

As demonstrated by \citet{helmi2000a}, the integral of motions space is a powerful tool to search for accreted satellites.
Indeed, before the merging process, Milky Way satellites have clumps in the integrals of motions space. The initial clumping should be present even after the satellite has completely mixed with the Galaxy and some systems survive as coherent structures for more than a Hubble time. 
 
Seven, possibly eight, Type~II GCs are clustered in a small region of the $L_{\rm PERP}$ vs.\,$L_{\rm Z}$ plane. 
 This evidence demonstrates that at least the seven Type~II GCs may be associated with a single accretion event.
\end{itemize}

In conclusion, our results show that similar MP properties are present in Milky Way and MC GCs, which also display a ChM consistent with the presence of both a 1G and a 2G group.
 The fact that the maximum helium variation and the fraction of 1G stars in MC and groups of Milky Way GCs with different origin, and possibly different parent galaxies, follow the same relation with cluster mass, suggest that GC mass is a universal parameter that determines the complexity of MPs in GCs.  

Evidence for a possible dependence of MP properties on the environment is provided by Type\,II GCs, which, using both our data and that of \citet{Wagner-Kaiser2017a}, are apparently not found in the SMC and LMC cluster populations.
seem to be absent either in the SMC and the LMC. In contrast, the possibility that at least seven out of thirteen known Type\,II GCs might be associated with a unique progenitor galaxy, possibly Gaia Enceladus, suggests that  their host galaxy has favoured the formation of these intriguing objects. 
 
\section*{acknowledgments} 
\small
We thank the anonymous referee for her/his contribution.
This work has received funding from the European Research Council (ERC) under the European Union's Horizon 2020 research innovation programme (Grant Agreement ERC-StG 2016, No 716082 'GALFOR', PI: Milone, http://progetti.dfa.unipd.it/GALFOR), and the European Union's Horizon 2020 research and innovation programme under the Marie Sklodowska-Curie (Grant Agreement No 797100). APM and MT acknowledge support from MIUR through the FARE project R164RM93XW SEMPLICE (PI: Milone).
CL acknowledges support from the one-hundred- talent project of Sun Yat-Sen University. HJ acknowledges support from the Australian Research Council through the Discovery Project DP150100862.

\bibliography{ms}
 
\end{document}